\documentclass[aps,prx,twocolumn,showpacs,floatfix,superscriptaddress]{revtex4}

\usepackage{times}
\usepackage{graphicx}
\usepackage{hyperref}
\hypersetup{colorlinks=true,linkcolor=blue}
\usepackage{amsmath}
\usepackage{wrapfig}
\usepackage{siunitx}

\usepackage{soul} 

\newcommand{\mean}[1]{\langle #1\rangle}

\begin{document}

\flushbottom
\title{Filtering Multiphoton Emission from State-of-the-Art Cavity QED}

\author{Carlos S\'anchez Mu\~noz }
\affiliation{CEMS, RIKEN, Wako-shi, Saitama, 351-0198, Japan}

\author{Fabrice P. Laussy }
\affiliation{Faculty of Science and Engineering, University of Wolverhampton,  Wulfruna St, WV1~1LY, United Kingdom}
\affiliation{Russian Quantum Center, Novaya 100, 143025 Skolkovo,  Moscow Region, Russia}

\author{Elena del Valle}
\affiliation{Departamento de
F\'isica Te\'orica de la Materia Condensada  and Condensed Matter Physics Center (IFIMAC), Facultad de Ciencias \\ Universidad
Aut\'onoma de Madrid, E-28049 Madrid, Spain}

\author{Carlos Tejedor}
\affiliation{Departamento de
F\'isica Te\'orica de la Materia Condensada  and Condensed Matter Physics Center (IFIMAC), Facultad de Ciencias \\ Universidad
Aut\'onoma de Madrid, E-28049 Madrid, Spain}

\author{Alejandro Gonz\'alez-Tudela}
\affiliation{Max--Planck Institut f\"ur Quantenoptik, 85748 Garching, Germany}

\newcommand{\ketbra}[2]{\ket{#1}\bra{#2}}
\newcommand{\ket}[1]{|#1\rangle}
 \newcommand{\bra}[1]{\langle #1|}
 \newcommand{\down}{\ket{g}\bra{e}}
\newcommand{\up}{\ket{e}\bra{g}}
\newcommand{\downd}{\ket{-}\bra{+}} 
\newcommand{\upd}{\ket{+}\bra{-}}
\newcommand{\app}{a^\dagger}
\newcommand{\ssp}{\sigma^\dagger}
\newcommand*{\Resize}[2]{\resizebox{#1}{!}{$#2$}}%

\begin{abstract}
Engineering multiphoton states is an outstanding challenge with applications in multiple fields, such as quantum metrology, quantum lithography or even biological systems. State-of-the-art methods to obtain them rely on post-selection, multi-level systems or Rydberg atomic ensembles. Recently, it was shown that a strongly
driven two-level system interacting with a detuned cavity mode can be engineered
to continuously emit $n$-photon states. In the present work, we show that spectral filtering of its emission relaxes considerably the requirements on the system parameters even to the more accessible \emph{bad-cavity} situation, opening up the possibility of implementing this protocol in a much wider landscape of different platforms.
This improvement is based on a key observation: in the imperfect case where only a certain fraction of emission is composed of $n$-photon states, these have a well defined energy separated from the rest of the signal, which allows to reveal and purify multiphoton emission just by frequency filtering. We demonstrate these results by obtaining analytical expressions for relevant figures of merit of multiphoton emission, such as the $n$-photon coupling rate between cavity and emitter, the fraction of light emitted as $n$-photon states, and $n$-photon emission rates. This allows us to make a systematic study of such figures of merit as a function of the system parameters and demonstrate the viability of the protocol in several relevant types of cavity QED setups, where we take into account the impact of their respective experimental limitations.
\end{abstract}
\date{\today} \maketitle

\section{Introduction \label{sec:intro}}

Non-classical states of light are a fundamental ingredient in the development of photonic quantum technologies, such as quantum communication~\cite{kimble08a}, quantum metrology~\cite{giovannetti04a}, lithography~\cite{dangelo01a}, spectroscopy~\cite{lopez15a,dorfman16a} or biological sensing~\cite{denk90a,horton13a}. The generation of single-photons is experimentally accessible by several methods, such as exploiting the single-photon non-linearity of natural and artificial atoms~\cite{xu99a,painter99a,kimble76a,matthiesen12a,proux15a,somaschi16a,arcari14a} or using correlated photon pairs in non-linear crystals~\cite{bruno14a} or biexciton states~\cite{callsen13a,dousse10a,muller14a}. However, obtaining multiphoton states is a much more challenging task as the $n$-photon interactions are generally very weak. Current methods to generate multiphoton states in the optical regime are mostly probabilistic and rely on using single-photons, linear optics and post-selection to build up higher photon numbers. Unfortunately, they suffer from an exponential 
scaling of success with 
increasing photon number~\cite{yao12a}. Thus, it is still of fundamental and practical interest to search for novel mechanisms which allow us to engineer light at $n$-photon level.

\begin{figure*}[t!!!]
\begin{center}
\includegraphics[width=0.95\textwidth]{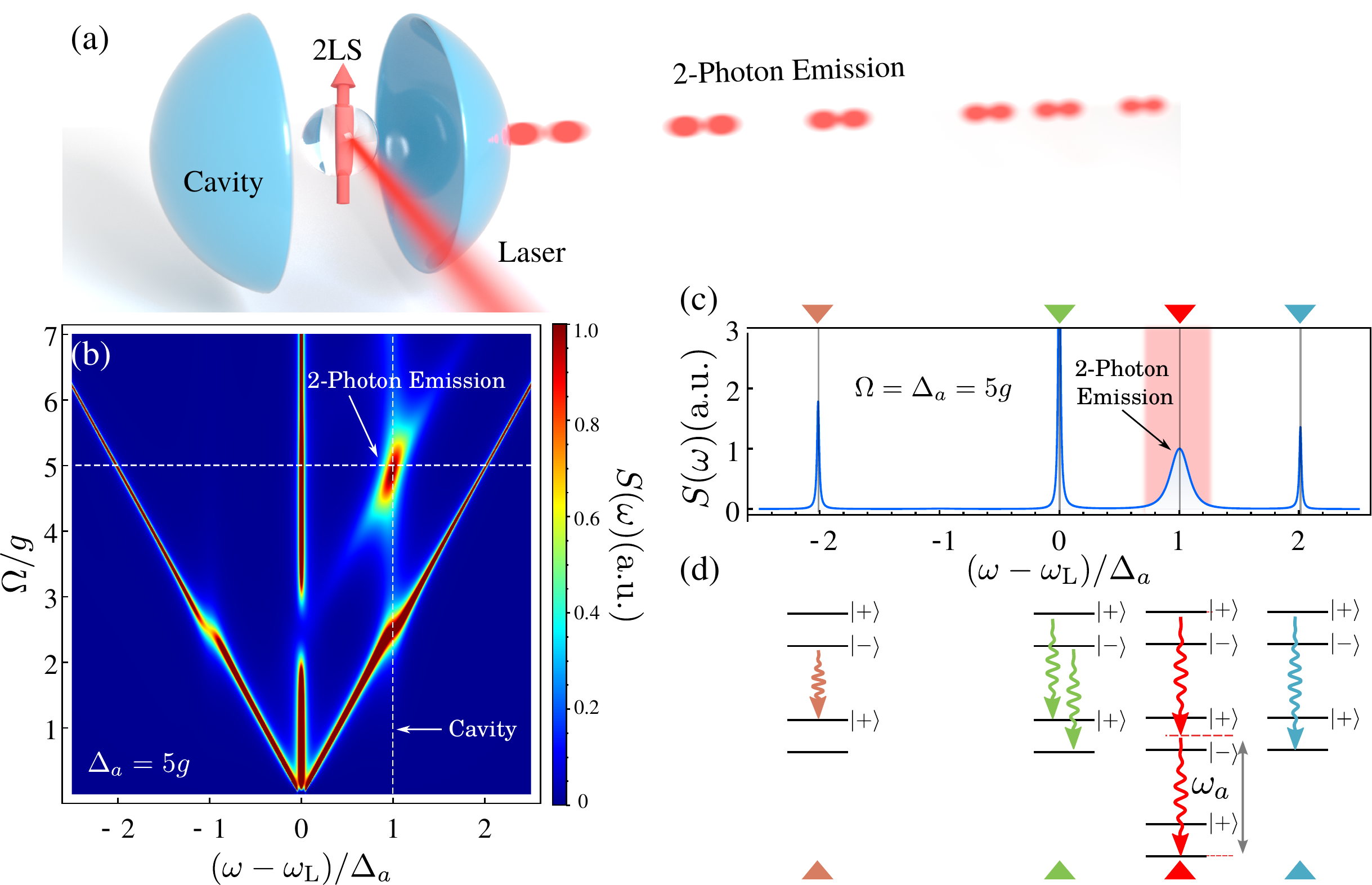}
\end{center}
\caption{(a) Scheme of the proposed setup: a two-level system is coupled to a cavity and strongly driven by a classical field. By selecting the proper cavity frequency, the system emits a continuous stream of photon pairs.
(b) Evidence of two-photon emission in the cavity fluorescence spectrum, plotted as a function of the driving field amplitude. The driving laser is in resonance with the 2LS, and the cavity is detuned by $\Delta_a = 5g$. Mollow sidebands appear in the spectrum at $\omega = \omega_\mathrm L \pm 2\Omega$. When the cavity frequency lies between a sideband and the central peak, $\Delta_a = \pm \Omega$, two-photon emission is seen as a clear feature in the spectrum. (c) Spectrum in the regime of two-photon emission, at $\Omega = \Delta_a$. (d) Transitions in the ladder of dressed states giving rise to the four peaks featured in panel (c).  Simulation parameters: $\gamma_a = 1.3 g$, $\gamma_\sigma = 0.01g$.  }
\label{fig1}
\end{figure*}

With this motivation, several methods have been proposed to generate $n$-photon states in very different scenarios, such as Rydberg atomic ensembles~\cite{bienias14a,maghrebi15a,fistenberg16a}, atoms coupled to waveguide systems~\cite{gonzaleztudela15b,douglas16a,gonzaleztudela17c}, using multilevel atoms~\cite{koshino13a,sanchezmunoz15a,chang16a,sanchezburillo16a,hargart16a} or cavity QED setups~\cite{law96a,sanchezmunoz14a}. Particularly appealing, due to its simplicity, is the proposal in Ref.~\cite{sanchezmunoz14a}, which requires a single coherently driven two-level system (2LS), with driving amplitude $\Omega$, coupled to a single cavity mode, with strength $g$. In the strong driving limit, i.e., $\Omega\gg g$, it was shown that by placing appropriately the cavity resonance, the system can be brought to emit $n$-photon states (termed $n$-photon bundles). However, the broadening introduced by the cavity and 2LS losses, denoted as $\gamma_{a/\sigma}$, also leads to the emission of photons that are not released in the 
form of $n$-photon bundles, and therefore, contaminate the $n$-photon character of the output. In order to obtain emission with a large fraction of $n$-photon states, which defines the \emph{purity} of $n$-photon emission~\cite{sanchezmunoz14a},
 the coupling rate between cavity and 2LS was taken in the strong coupling regime with $g > \gamma_{a,\sigma}$.

Here, we revisit this proposal to obtain analytical expressions for the figures of merit of the mechanism, such as the $n$-photon couplings, efficiency rates or the \emph{purity} of $n$-photon emission. Moreover, we identify the spectral distribution of both the spurious and the $n$-photon component of the cavity emission. In particular, we show that they are emitted in different frequency windows, such that the contaminating photons can be mostly suppressed with appropriate frequency filtering. This simple insight leads to better figures of merit for the $n$-photon emission and lifts the strong coupling requirements to observe such effect. In the particular case of two-photon emission, we show how high purities can be obtained as long as the rate of emission into the cavity with respect to free space, i.e., the cooperativity $C=4g^2/(\gamma_a\gamma_\sigma)$, is large (the so-called \emph{bad-cavity} limit). This opens up the possibility of implementing this proposal in a wide variety of platforms 
such as semiconductor QDs~\cite{laucht12a,hennessy07a,lodahl15a, makhonin14a} or atom-nanophotonics~\cite{goban13a,thompson13a} where the bad-cavity limit has been observed, but the strong coupling condition is hard to obtain. Another consequence of the spectral isolation of $n$-photon bundles is that they manifest as a clear feature in the spectrum, offering the simplest smoking gun to experimentally evidence $n$-photon emission (see Fig.~\ref{fig1}).

The outline of the paper reads as follows: in Section~\ref{sec:system} we review the setup and mechanism proposed in Ref.~\cite{sanchezmunoz14a} and derive analytical expressions for the figures of merit of $n$-photon emission, allowing us to show clearly how these depend on the system parameters. Then, in Section~\ref{sec:filter}, we show  the spectral distribution of both the cavity emission in the $n$-photon resonance configuration and characterize the improvement in the figures of merit when filtering out the uncorrelated photons. In Section~\ref{sec:setups}, we analyze several experimental limitations of the platforms where our proposal can be implemented, such as semiconductors  or atomic cQED, and study their impact on the figures of merit for $n$-photon emission. Finally, we summarize our findings and point to future work in Section~\ref{sec:conclu}.

\section{System and general mechanism \label{sec:system}}

The proposal in Ref.~\cite{sanchezmunoz14a} is based on two cornerstones of Quantum Optics, namely, resonance fluorescence~\cite{mollow69a} and cavity QED~\cite{haroche89a,miller05a}. In particular, it requires a single 2LS interacting both with a classical field, with amplitude $\Omega$, and a quantum one, most simply implemented by a single-mode cavity coupled to the 2LS with strength $g$ (see Fig.~\ref{fig1}(a)). For simplicity, we assume that the laser frequency, $\omega_L$, is resonant with the transition frequency of the 2LS, $\omega_\sigma$, as it makes the analysis simpler, at the expense of only weakly worsening the efficiency. 
Besides, the system is operated in the dispersive regime where the 2LS and cavity are detuned.
In this configuration, strong correlations can be sustained between them even if their coupling is closer to a weak-coupling scenario with a small admixture of the bare states~\cite{laussy12e}. In these conditions, the total Hamiltonian of the combined system reads (using $\hbar=1$):
\begin{equation}
\label{eq:Ham}
 H=\Delta_a a^\dagger a+g(a^\dagger\sigma +\sigma^\dagger a)+\Omega(\sigma+\sigma^\dagger)\,,
\end{equation}
which we have written in a frame rotating with the laser/2LS frequency to make it time independent, such that $\Delta_a=\omega_a-\omega_\sigma$ is the cavity-2LS detuning. The operators $a^\dagger/a$ ($\sigma^\dagger/\sigma$) represent the creation/annihilation operators of the cavity (2LS) system. To take into account the cavity and 2LS losses, the Hamiltonian picture must be upgraded to a master equation description which takes into account the coupling to external baths. This results in an effective master equation which can be written:
\begin{equation}
\frac{d\rho}{dt} = -i\left[H,\rho \right] + \frac{\gamma_{\sigma}}{2}\mathcal{L}_{\sigma}[\rho] + \frac{\gamma_a}{2}\mathcal{L}_{a} [\rho]\, ,
\label{eq:master-equation}
\end{equation}
where $\mathcal{L}_O [\rho] \equiv 2 O\rho O^\dagger - O^\dagger O \rho - \rho O^\dagger O $ is a Lindblad term describing the leakage of photons from both the cavity and the bare 2LS.

When $g\equiv 0$, we recover the standard resonance fluorescence situation. This configuration has been traditionally exploited in the weak driving limit, $\Omega\ll\gamma_\sigma$, to generate individual single-photons~\cite{kimble76a,matthiesen12a,proux15a,somaschi16a} emitted within a small spectral windows around the 2LS frequency. In the strong driving limit, $\Omega\gg\gamma_\sigma$, the classical field dresses the 2LS levels, leading to two new eigenstates of the system, $\ket{\pm}=\frac{1}{\sqrt{2}}\left(\ket{g}\pm \ket{e}\right)$, with eigenenergies $E_{\pm}=\omega_\sigma \pm \Omega$. These dressed states deform the incoherent spectrum of 2LS from a single peak to the well-known Mollow triplet~\cite{mollow69a}. This spectral shape can be easily understood from single photon transitions in the dressed state picture~\cite{cohentannoudji77a} [see Fig.~\ref{fig1}(d)], corresponding to $\ket{\pm}\rightarrow\ket{\pm}$ and $\ket{\mp}\rightarrow \ket{\pm}$, which explains why three peaks appear at 
frequencies $\omega_\sigma (\pm 2\Omega)$. Something that generally goes unnoticed is that, in the strong driving limit, the 2LS also provides non-linearities 
at the multiphoton level. These non-linearities can be understood as $n$-photon transitions in the ladder of dressed states, going from a state $|\pm\rangle$ to a state $|\mp\rangle$, $n$ rungs below. As illustrated for the case $n=2$ in Fig.~\ref{fig1}(d), these processes are out of resonance from the first-order transitions associated to the Mollow triplet, and for the case of photons of equal energy it is easy to see that their energy is $\pm 2\Omega/n$. These multiphoton transitions are typically hidden by the single-photon processes, as they are of smaller order. However, they can be made visible, e.g., using frequency resolved correlations~\cite{delvalle12a,gonzaleztudela13c} as recently experimentally observed \cite{peiris15a,peiris17a}. Remarkably, these correlations in frequency space are strong enough to violate classical inequalities~~\cite{sanchezmunoz14b,peiris15a}.

Unfortunately, the emission of these strongly correlated photons is scarce, and therefore, difficult to exploit. An alternative and powerful way of capitalizing on these photons consists of coupling the strongly driven 2LS to a cavity (making $g\neq 0$ in our model), and setting the cavity energy exactly at these multiphoton resonances, $\Delta_a^{(n)}=\pm 2\Omega/n$. Using that simple prescription, it was shown by using a quantum jump simulation~\cite{sanchezmunoz14a} that one is able to Purcell-enhance the $n$-th photon process with respect to the single photon ones, leading to (almost) perfect emission in groups or \emph{bundles} of photons.

Fig.~\ref{fig1} illustrates this effect for the case of ${n=2}$ photons. In particular, in panel (b) we show the evolution of the incoherent part of the cavity spectrum, ${S(\omega)\propto \lim_{t\rightarrow\infty}\int d\tau \mean{a^\dagger(t)a(t+\tau)}e^{i\omega t}}$, with increasing driving $\Omega$ for a fixed cavity-2LS detuning of $\Delta_a=5g$ and $\gamma_a=1.3g$ and $\gamma_\sigma=0.01$, such that $C\sim 300$. These parameters are similar to those reported in experimental works that have already performed measurements similar to what we show in Fig.~\ref{fig1}(b)~\cite{kim14a,fischer16a}.  The main difference between those measurements and what we show here lies in the range of $\Omega$ considered, which in such experiments was not taken to the values where the new effects that we report clearly manifest themselves.
We can distinguish three different regimes:
\begin{itemize}
 \item When the driving is small, we only observe a single peak at the 2LS frequency. The main effect of the cavity here is to renormalize the broadening of such a peak to $\sim \gamma_\sigma+ \frac{4 g^2\gamma_a}{4\Delta_a^2+\gamma_a^2}$, that is, the sum of the individual 2LS decay plus the broadening provided by the cavity emission.
 
 \item As the driving increases, the cavity spectrum $S(\omega)$ starts to display a triplet structure, that is distorted by the finite detuning, $\Delta_a$, as it breaks the symmetry between the upper/lower sidebands from the bare Mollow triplet. Interestingly, when $\Omega$ is such that $\Delta_a$ is resonant with the upper sideband, i.e., $\Delta_a^{(1)}=2\Omega=5g$, one observes a complete depletion of the central peak, and a very strong asymmetry between the two sidebands. The physics at hand occurs at the single photon level.
 
 \item The most interesting regime for this manuscript occurs when $\Omega$ is such $\Delta_a=\Delta_a^{(2)}=\Omega=5g$. Around this value of $\Omega$, we observe in Fig.~\ref{fig1}(b) the appearance of an extra peak in the cavity spectrum which is precisely the one that generate the photon pairs predicted in Ref.~\cite{sanchezmunoz14a}. To make it more clear, we plot an horizontal cut at $\Omega=5g$ in Fig.~\ref{fig1}(c). The four peaks can be easily identified with processes in the dressed state picture, as schematically depicted in Fig.~\ref{fig1}(d): on the one hand, we observe the conventional single-photon processes at $\omega_\sigma-2\Omega,\omega_\sigma$ and $\omega_\sigma+2\Omega$; on the other hand, at the cavity frequency, between the central peak and the upper sideband, we observe a peak that corresponds to the cascaded emission of two-photons at frequencies $\omega_\sigma+\Omega$. This peak is broader than the other due to the Purcell enhancement by the cavity.
 \end{itemize}

 To certify that one is dealing with true $n$-photon emission, one can study the statistics of the output field, as shown in Refs.~\cite{sanchezmunoz14a,sanchezmunoz15a,chang16a}. However, in this manuscript we will use the appearance of these extra peaks in $S(\omega)$ as the signature for $n$-photon emission, as it is the simplest experimentally relevant smoking gun for multiphoton emission. A similar method has been used, for instance, to experimentally evidence Purcell-enhanced two photon emission in a biexcitonic radiative cascade~\cite{ota11a}.

\subsection{Analytical derivation of the $n$-photon coupling rate}

After having illustrated the mechanism for a particular situation, one of the main goals of this manuscript is to gain analytical understanding on the figures of merit of the emission. As shown in Fig.~\ref{fig1}(c), all the non-linear processes appearing in $S(\omega)$ are well understood in the dressed state picture. Thus, it is enlightening to write the Hamiltonian in the dressed state basis where it reads:
\begin{equation}
H = \Omega\, \tilde\sigma_z +  \Delta_a \,\app a + \frac{g}{2}\left\{ \app(\tilde\sigma -\tilde{\sigma}^\dagger + \, \tilde\sigma_z) + \mathrm{h.c.}\right\}\,,
\label{eq:hamiltonian-dressed}
\end{equation}
where $\tilde\sigma =\ket{-}\bra{+}$ and $\tilde\sigma_z=\ketbra{+}{+}-\ketbra{-}{-}$. In the strong driving limit, $\Omega\gg g$, and when the cavity is close to the $n$-th photon resonance, i.e., $\Delta_a \approx \Delta_a^{(n)}$, the energy levels of the Hamiltonian are structured in manifolds $\mathcal{E}_{m,n} = \{|+,m\rangle, |-,m+n\rangle \}$ (where $m$ is the number of photons in the cavity) such that the energy separation between levels inside a manifold is much smaller than the energy separation between different manifolds. This energy separation allows one to perform an adiabatic elimination of the fast degrees of freedom to construct an effective Hamiltonian that does not couple the manifolds between them. Under these conditions, the dynamics can then be described by an effective $n$-photon coupling Hamiltonian:
\begin{equation}
H_\mathrm{eff}^{(n)} =  \Omega \tilde{\sigma}_z + \Delta_a a^\dagger a + g^{(n)}\left(\tilde{\sigma}^\dagger a^n + \tilde\sigma {a^\dagger}^n \right) \, ,
\label{eq:n-photon-hamiltonian}
\end{equation}
which generates $n$-photon Rabi oscillations between the states $|+,m\rangle$ and $|-,m+n\rangle$ with rate (see Appendix~\ref{appendix:n-photon}):
\begin{equation}
g^{(n)} = \frac{g^n}{2 (n-1)!^2}\left(\frac{n^2}{4\Omega}\right)^{n-1}\,.
\label{eq:gn}
\end{equation}
There are also small renormalizations of the bare 2LS and cavity energies that shift the $n$-photon resonance from the values $\Delta_a^{(n)} = \pm 2R/n$. For simplicity, we will omit writing these renormalizations in the following descriptions, but include them consistently to perform the calculations.  

Performing the same change of basis into the 2LS Lindblad operators, we arrive to:
\begin{equation}
\frac{\gamma_{\sigma}}{2}\mathcal{L}_{\sigma}[\rho] \approx \left(\frac{\gamma_{\sigma}}{8} \mathcal{L}_{\tilde\sigma}+ \frac{\gamma_{\sigma}}{8} \mathcal{L}_{\tilde{\sigma}^\dagger} + \frac{\gamma_{\sigma}}{2}\, \mathcal{L}_{\tilde{\sigma}^\dagger \tilde{\sigma}}\right)[\rho] \, ,
\label{eq:lindblad-effective}
\end{equation}
where other fast rotating terms are eliminated under the assumption $\Omega \gg \gamma_{\sigma}$. Therefore, the bare 2LS decay transforms in the dressed state basis into: i) an effective decay rate $\gamma_{\tilde\sigma}=\frac{\gamma_\sigma}{4}$; ii) an  effective pumping $P_{\tilde\sigma}=\frac{\gamma_\sigma}{4}$; and ii) a dephasing term with rate $\gamma_{\tilde{\phi}}=\gamma_\sigma$.

By solving the master equation with the $n$-photon Hamiltonian in the dressed basis,
\begin{multline}
\frac{d\rho}{dt} = -i\left[H^{(n)}_\mathrm{eff},\rho \right]+\frac{\gamma_a}{2}\mathcal L_a [\rho]  +\frac{\gamma_{\tilde\sigma}}{2} \mathcal{L}_{\tilde\sigma}[\rho]+\\ \frac{P_{\tilde\sigma}}{2} \mathcal{L}_{\tilde{\sigma}^\dagger}[\rho] +\frac{\gamma_{\tilde\phi}}{2} \, \mathcal{L}_{\tilde{\sigma}^\dagger \tilde{\sigma}}[\rho],
\label{eq:n-photon-ME}
\end{multline}
we calculate the amount of cavity population introduced via the $n$-photon coupling of Eq.~\eqref{eq:n-photon-hamiltonian} under the assumption ${g^{(n)}\ll\gamma_a}$, obtaining (see Appendix~\ref{appendix:na_n}):
\begin{equation}
n_a^{(n)}\approx n^2\frac{\kappa^{(n)}}{\Gamma_{\tilde\sigma}(n \gamma_a +\Gamma_{\tilde\sigma} + \gamma_{\tilde\phi})+ \kappa^{(n)} n\gamma_a}P_{\tilde{\sigma}}\,,
\label{eq:na-nphoton}
\end{equation}
where $\Gamma_{\tilde\sigma}=\gamma_{\tilde\sigma}+P_{\tilde\sigma}$, and $\kappa^{(n)}=\frac{4(n-1)!(g^{(n)})^2}{\gamma_a}$ is a generalized $n$-photon Purcell rate. In the limit $n\gamma_a\gg\gamma_\sigma$, the previous expression simplifies to:
\begin{equation}
n_a^{(n)}\approx n\frac{\kappa^{(n)}}{\gamma_a}\frac{P_{\tilde\sigma}}{\kappa^{(n)}+\Gamma_{\tilde\sigma}}\,,
\label{eq:na-nphotonsimp}
\end{equation}
in which each term has a transparent meaning: i) the term $\frac{P_{\tilde\sigma}}{\kappa^{(n)}+\Gamma_{\tilde\sigma}}$ is the population of the dressed 2LS that mediates the $n$-photon coupling; ii) the term $\frac{\kappa^{(n)}}{\gamma_a}$ is the ratio between the effective pumping to the cavity mode through the $n$-photon Purcell decay rate, $\kappa^{(n)}$, and the cavity decay rate, $\gamma_a$; iii) the factor $n$ takes into account that every time a jump occurs in the effective dressed 2LS, $n$ photons are introduced in the cavity. One can confirm, by looking at the output field of the cavity, that these photons introduced through the cavity in groups of $n$, are also emitted as such as was confirmed in Ref.~\cite{sanchezmunoz14a,sanchezmunoz15a}. The timescale between the $n$-photon states is $\sim (\kappa^{(n)}+\gamma_{\bar{\sigma}})^{-1}$, as it is the time it takes to reload the dressed 2LS after an $n$-photon quantum jump. The spectral width of the $n$-photon wavepacket is of the 
order of $n\gamma_a$. Thus, in order to obtain antibunched $n$-photon emission it must be satisfied that~\cite{chang16a} $n\gamma_a\gg \kappa^{(n)}+\gamma_{\tilde{\sigma}}$.
\begin{figure}[t!!!]
\begin{center}
\includegraphics[width=0.99\columnwidth]{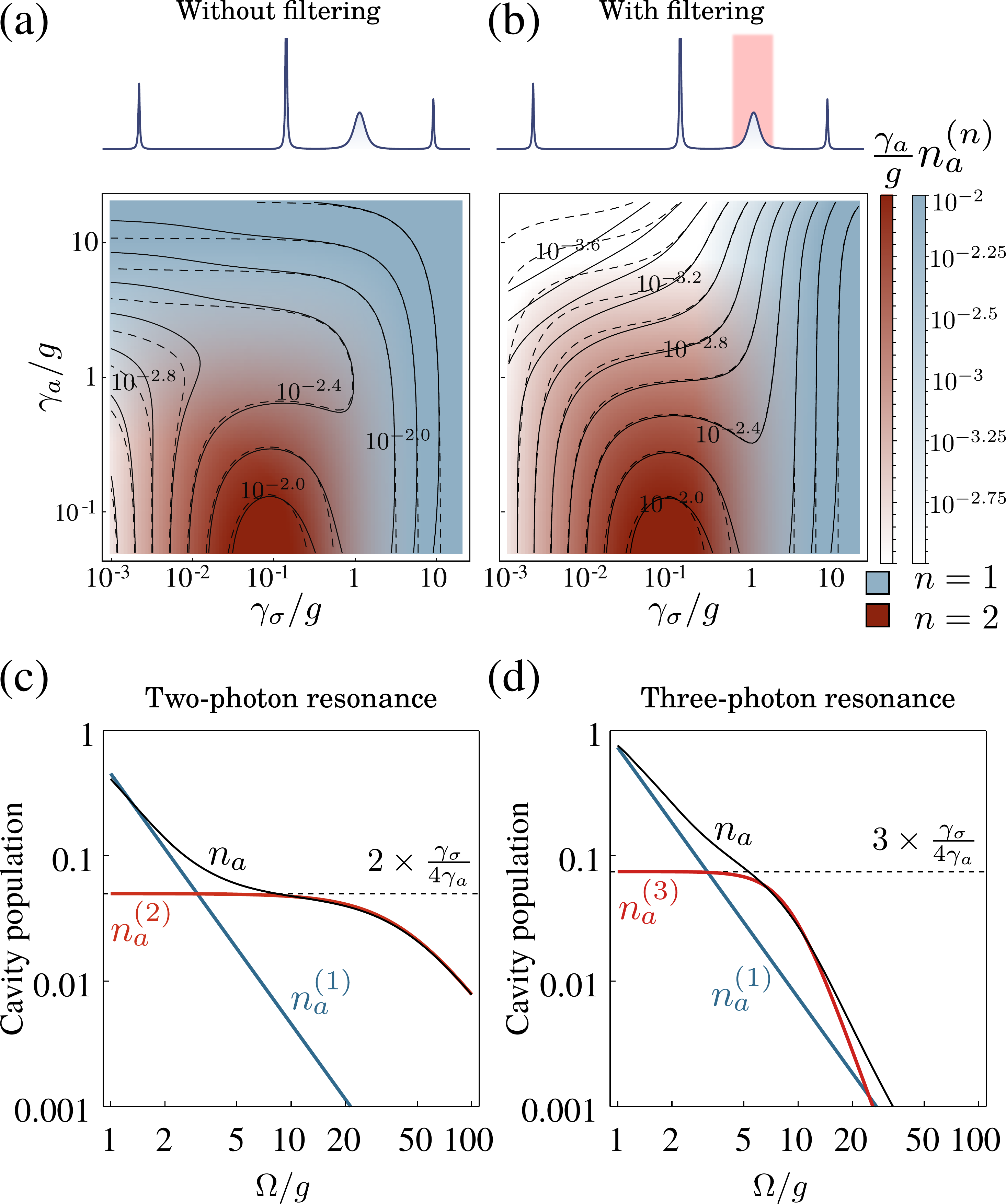}
\end{center}
\caption{ (a) Joint plot of two-photon [$n_a^{(2)}$, red] and one-photon [$n_a^{(1)}$, blue] emission rates as a function of cavity ($\gamma_a$) and 2LS ($\gamma_\sigma$) decay rates. 
 Contour lines correspond to the total population $n_a$ computed numerically (solid) and analytically (dashed), based on the assumption $n_a \approx n_a^{(2)} + n_a^{(1)}$. The good agreement confirms the validity of such assumption and of Eqs.~\eqref{eq:na-nphoton} and \eqref{eq:na-1photon-expansion}. (b) Two-photon emission rate [$n_a^{(2)}$, red] and one-photon emission rate at the cavity frequency [$n_{a,\mathrm f}^{(1)}$, blue]. Contour lines correspond to the total filtered emission $n_{a,\mathrm f}$, computed numerically (solid) and analytically, based on the assumption $n_{a,\mathrm f}\approx n_a^{(2)}+n_{a,\mathrm f}^{(1)}$ and Eqs.~\eqref{eq:na-nphoton} and \eqref{eq:n1filt}. $\Omega = 20g$ in panels (a) and (b).
{(c)} Components of the cavity population as a function of $\Omega$ when the cavity is at the two-photon resonance ($\Delta_a = \Omega$).
 Black: total cavity population. Blue: Single-photon component of cavity population as given in Eq.~\eqref{eq:na-1photon-expansion}. Red: two-photon component of cavity population $n_a^{(2)}$ as given by Eq.~\eqref{eq:na-nphoton}.  Dashed, horizontal lines mark the weak driving limit of $n_a^{(n)}$ given by Eq.~\eqref{eq:na-n-lowpumping}. {(d)} Same as in (c), for the three-photon resonance. Simulation parameters, $\gamma_a = 0.1g$ and $\gamma_\sigma=0.01g$}
\label{fig2}
\end{figure}

However, it was shown~\cite{sanchezmunoz14a} that even in the perfect resonant case the $n$-photon emission at the output of the cavity field is always contaminated by the emission of uncorrelated single-photon states. The origin of such photons is that, when writing the effective Hamiltonian of Eq.~\eqref{eq:n-photon-hamiltonian}, we are neglecting the off-resonant, but first order, processes that also generate population in the cavity, which we denote by $n_a^{(1)}$. These are specially relevant when the broadening introduced by $\gamma_{a,\sigma}$ is considerable. Interestingly, this off-resonant population $n_a^{(1)}$ can also be analytically estimated in the limit $\Omega\gg g$. For example, for the $2$-photon resonant situation that we will be focusing on along this manuscript, $\Delta_a=\Delta_a^{(2)}$, and in the limit of $\gamma_\sigma\ll \Omega,\Delta_a,\gamma_a$, it can be approximated by ({see Appendix~\ref{appendix:na_n}):

\begin{small}
\begin{equation}
n_{a}^{(1)}\approx \frac{2g^2(\gamma_a^2+28\Omega^2)}{(\gamma_a^2+4\Omega^2)(\gamma_a^2+36\Omega^2)}
 + \frac{32g^2\Omega^2(\gamma_a^4+432\Omega^4)\gamma_{\tilde\sigma}}{\gamma_a(\gamma_a^2+4\Omega^2)^2(\gamma_a^2+36\Omega^2)^2}\,.
\label{eq:na-1photon-expansion}
\end{equation}
\end{small}

The more general expression as a function of $\Delta_a$ can be found in Appendix~\ref{appendix:na_n}. Finally, we make the assumption that the total population in the cavity is given by the sum of the two mechanisms described above: $n_a \approx n_a^{(1)} + n_a^{(n)}$. We numerically confirmed this is a good assumption in the regime of validity of the approximations.

\subsection{Characterizing $n$-photon emission}

Thanks to the analytical results that we developed in the previous Section, we can make a more systematic analysis of the main figures of merit of the $n$-photon emission of our proposal, namely, i) the $n$-photon emission rate through the cavity mode, that is given by $\gamma_a n_a^{(n)}$; and ii) the \emph{purity of $n$-photon emission}, defined as the fraction of the total population which given by $n$-photon population:
\begin{equation}
 \label{eq:purity}
 \pi_n= \frac{n_a^{(n)}}{n_a} \approx \frac{n_a^{(n)}}{n_a^{(1)}+n_a^{(n)}}\,.
\end{equation}

We start by showing a contour plot in Fig.~\ref{fig2}(a) of the single (in blue) and two-photon (in red) emission rates as a function of $(\gamma_a,\gamma_\sigma)/g$ for a situation with $\Omega=20g$ and $\Delta_a$ tuned to the $2$-photon resonance, i.e.,  $\Delta=\Delta_a^{(2)}=\Omega$. Both plots are superimposed to facilitate the comparison. We confirm that, in order for the two-photon process to dominate, it is required that $\gamma_{a,\sigma}<g$, which corresponds to the red region in the lower part of the figure. This was the regime considered in Ref.~\cite{sanchezmunoz14a}. As $\gamma_a$ increases, the importance of the $n$-photon emission rate also decreases as can be expected from the analytical formula of $n_a^{(n)}$ in Eq.~\eqref{eq:na-nphotonsimp} because $\kappa^{(n)}\propto 1/\gamma_a$. The behaviour with $\gamma_\sigma$ is, however, less trivial, as there is a trade-off between increasing the pumping of the effective two-level system, $P_{\tilde\sigma}\propto\gamma_\sigma$, but at the same time increasing its losses $\gamma_{\tilde\sigma}$, which results in an 
optimal $\gamma_\sigma$ as shown in Fig.~\ref{fig2}(a). As we have the analytical expressions for $n_a^{(n)}$, we can find the optimal $\gamma_\sigma^\mathrm{n,opt}$ which gives the largest amount of $n_a^{(n)}$ for a fixed set of parameters. The general expression for any $n$ reads
\begin{equation}
\gamma_{\sigma}^{(n,\mathrm{opt})} = \left(\frac{g n^2}{4\Omega}\right)^n\frac{8\Omega}{\sqrt{3 (n!)^3}}\,.
\end{equation}
For the case $n=2$ that we plot in Fig.~\ref{fig2}(a), the optimal decay is obtained when $\gamma_{\sigma}^{(2,\mathrm{opt})}=2\sqrt{2/3}g^2/\Omega\approx 0.08g$, which coincides with the numerical results shown in Fig.~\ref{fig2}(a).

Another interesting behaviour to explore is the dependence of the efficiencies with $\Omega$, as it is an experimentally tunable parameter. In Figs.~\ref{fig2}(c-d) we plot the dependence with $\Omega$ of the total cavity population (black), and its single (blue) and $n$-photon components (red) given by Eqs.~\eqref{eq:na-1photon-expansion} and \eqref{eq:na-nphoton}, for a system with $\Delta_a=\Delta_a^{(2-3)}$ fixed at the $2-3$ photon resonance respectively. This means that we change $\Delta_a$ accordingly as the driving increases.  There are several interesting conclusions to be extracted from this figure:
\begin{itemize}
 \item In both the two- and three-photon resonance situations the analytical formulas capture very well, both qualitatively and quantitatively, the full numerical results in all the regimes of $\Omega$ depicted in Figs.~\ref{fig2}(c-d).
 
 \item As expected at weak drivings, the single-photon population, $n_a^{(1)}$, dominates over the $n$-photon one as the detuning is not enough to supress one-photon processes. From the formula of $n_a^{(n)}$ one can actually obtain a simple formula for the limit of very weak driving:
 \begin{align}
n_a^{(n) }  &\approx  \frac{n}{4}\frac{\gamma_{\sigma}}{\gamma_a}, \quad (\Omega \rightarrow 0)\, .
\label{eq:na-n-lowpumping}
\end{align}
which is the horizontal line that we plot in Figs.~\ref{fig2}(c-d).
\item As the driving increases both $n_a^{(1)}$ and $n_{a}^{(n)}$ (and therefore $n_a$) decrease, but with a significant difference between the two and three-photon situation. While at the two-photon resonance case, $n_a^{(1)}$ and $n_a^{(2)}$ decrease with the same scaling with $\Omega$, which allows $n_a^{(2)}$ to dominate even for very large $\Omega$; in the case of three photons, $n_a^{(3)}$ decreases faster than $n_a^{(1)}$ such that there will be an optimal $\Omega$ to maximize $n_a^{(3)}$ over $n_a^{(1)}$. This behaviour can also be explained from the analytical formulas by expanding them in the $\Omega\rightarrow \infty$ limit:
\begin{align}
\label{eq:nastrongOm}
n_a^{(n)}\approx & \frac{g^{2n}}{\Omega^{2(n-1)}} A_n\propto \frac{1}{\Omega^{2(n-1)}}, \\
 n_a^{(1)}\approx & \left(\frac{g^2}{\Omega^2}\right)\frac{n^3(2+n^2)\gamma_{\sigma}+n(n^4-n^2+2)\gamma_a}{16(n^2-1)^2\gamma_a}\nonumber\\\propto &\frac{1}{\Omega^2}\,,
\end{align}
where $A_n$ reads:
\begin{equation}
A_n = \left[16^{n-1}\gamma_a(2n\gamma_a+3\gamma_\sigma)n^{2(1-2n)}(n-1)!^3 \right]^{-1}.
\label{eq:c6-An}
\end{equation}

The convergence to that limit is obtained when $(\Omega/g)^{2(n-1)}\gg C$ [$\Omega\gg\gamma_{a,\sigma}$] for $n_a^{(n)}$ [$n_a^{(1)}$].

\end{itemize}

From this discussion, we observe there is a non-trivial trade-off between absolute multiphoton emission rates and the \emph{purity} of the multiphoton source as defined in Eq.~\eqref{eq:purity}. Using the analytical expressions we developed, we can find an approximated formula for the $\pi_n$ which reads:
\begin{multline}
\pi_n \approx \left\{1+ \phantom{\frac{4g(\gamma_a^2 + 28\Omega^2)}{(\gamma_a^2+4\Omega^2)(\gamma_a^2+36\Omega^2)}} \right. \\ \left. \left[ \frac{4g(\gamma_a^2 + 28\Omega^2)}{(\gamma_a^2+4\Omega^2)(\gamma_a^2+36\Omega^2)}+ \frac{32g^2\Omega^2(\gamma_a^4+432\Omega^4)\gamma_\sigma}{\gamma_a(\gamma_a^2+4\Omega^2)^2(\gamma_a^2+36\Omega^2)^2}  \right] \right. \\
\left. \times \left[\frac{4\gamma_a}{n\gamma_{\tilde\sigma}} +\frac{ \Omega^{2(n-1)}}{g^{2n} A_n}  \right] \right\}^{-1} \, .
\label{eq:c6-purity-analytical-formula}
\end{multline}

We numerically checked (not shown) that the $\pi_n$ obtained from the analytical formulas agree quantitative well with the ones that can be obtained from the photon counting distribution of the output field~\footnote{Notice that the definition of $\pi_n$ introduced in Ref.~\cite{sanchezmunoz14a} differs slightly with the one used here. In that work, it was defined as $\pi_n = \frac{\lambda_n}{\lambda_n + \lambda_1}$, where $\lambda_n$ are fitted from photon counting distributions and correspond to  $\lambda_n = \gamma_a n_a^{(n)}/n$. The comparison was made adopting the definition used in this text and defining $\pi_n=\frac{\lambda_n/n}{\lambda_n/n+\lambda_1}$. }, as was introduced in Refs.~\cite{sanchezmunoz14a,sanchezmunoz15a}.

Equation~\eqref{eq:c6-purity-analytical-formula} allows us to obtain simple asymptotic expressions. For example, in the large driving limit, $\Omega\rightarrow \infty$, one obtains the following formula for the purity of two-photon emission:
\begin{equation}
\pi_2 \approx \left(1 + \frac{7}{18}\frac{\gamma_a^2}{g^2}+ \frac{8}{3C} + \frac{21}{18 C} + \frac{8g^2}{\gamma_a^2 C^2} \right)^{-1},
\label{eq:pi2}
\end{equation}
where the term proportional to $(\gamma_a/g)^2$ evidences that one needs to have strong coupling parameters, $\gamma_a \ll g$, in order to obtain values of $\pi_2$ close to one, as was already observed in Ref.~\cite{sanchezmunoz14a}. For values of $n$ larger than 2, the asymptotic expression for the purity is $\pi_{n>2}\propto \frac{1}{\Omega^{2(n-1)}}$. This tells us that for $n=2$ case, one can always purify the two-photon emission by going to larger drivings, whereas for the $n>2$ case one has to find the optimal driving that maximizes the purity for a given set of parameters. The explanation for this is that $n_a^{(n)}$ is given, for $n=1$, by an off-resonant, first-order process, and for $n>1$, by a resonant, $n$-th order process. Only in the case of $n=2$ these two properties (on resonance and second-order) compensate and yield values of $n_a^{(2)}$ larger than $n_a^{(1)}$ at large drivings. Though, in principle it is possible to find an optimal $\Omega$ for $n>2$ by differentiating the expression of $\pi_n$, its expression its too cumbersome to write here. An alternative way to obtain an approximated optimal $\Omega$ consists of finding the point where the asymptotic expression of $n_a^{(n)}$ at small and large driving cross. The latter procedure leads to a simple expression for the optimal driving, which reads:
\begin{equation}
\Omega^{(n)}_\mathrm{opt} \approx \left( \frac{4 \gamma_a g^{2 n}A_n}{n \gamma_{\sigma}}\right)^{\frac{1}{2(n-1)}} \, .
\end{equation}

We checked numerically that this is indeed a very good approximation by comparing with the exact numerical results.\\[10pt]

\section{Improvement by frequency filtering \label{sec:filter}}

\subsection{Spectral distribution}
As we have seen in Fig.~\ref{fig1}, the multiphoton emission in our setup manifests as an extra peak in the incoherent cavity spectrum. However, together with this extra peak, we still observe three extra peaks reminiscent from the single-photon transitions of the Mollow triplet. A key observation is that the photons at these frequencies are the main origin of the small fraction of spurious photons that contaminate the multiphoton emission, that we labelled as $n_a^{(1)}$. The fact that these photons appear at frequencies well separated from the multiphoton peak allows one to purify the multiphoton source by frequency filtering, which as already been proven to be a way to optimize photon correlations in other situations~\cite{gonzaleztudela15a,delvalle13a}. Before analysing the effect of the filtering, let us first show how to estimate the fraction of the total emission corresponding to each of the spectral peaks. Formally, the incoherent part of the cavity spectrum can be calculated in terms of the eigenvalues, $\lambda_\beta$, and eigenvectors of the Liouvillian of the system as follows~\cite{delvalle11a,delvalle_book10a}:
\begin{multline}
S(\omega) = \frac{1}{\pi} \sum_\beta\left[\frac{ (\gamma_\beta/2)  L_\beta}{(\omega-\omega_\beta)^2+(\gamma_\beta/2)^2} \right.
\\
-\left.\frac{(\omega-\omega_\beta)K_\beta}{(\omega-\omega_\beta)^2+(\gamma_\beta/2)^2}\right]\,,
\label{eq:c6-spectrum-lorentzians}
\end{multline}
that is, a sum of Lorentzians centered at $\omega_\beta =\Im\{\lambda_\beta\}$ with linewidth $\gamma_\beta= 2\Re\{\lambda_\beta\}$ (plus a dispersive part which takes into account possible interference between them).  The weights $L_\beta,K_\beta$ are obtained from a combination of the eigenvectors of the Liouvillian and the steady-state of $\rho$. It can be trivially shown that $L_\beta$ satisfy that $n_a=\sum_\beta L_\beta$.  Using that information, it is possible to estimate the amount of cavity population emitted at the frequency of the cavity, $n_{a,\mathrm{f}}$, by summing those $L_\beta$ whose corresponding $\omega_\beta$ is close to the cavity frequency:
\begin{align}
 \label{eq:L}
n_{a,\mathrm{f}}=\sum_{\omega_\beta \approx \omega_a} L_\beta\,.
\end{align}

We can now estimate the amount of light from the single-photon processes emitted at the cavity frequency, $n_{a,\mathrm f}^{(1)}$, which is the one that we will not be able to get rid of by frequency filtering. This is done by using a similar analysis, but truncating the cavity's Hilbert space to one photon in order to exclude $n$-photon processes from the dynamics ({see Appendix~\ref{appendix:n1-filt}}). The approximated expression for $n_{a,\mathrm f}^{(1)}$ reads, in the limit of $\Omega \gg g,\gamma_a,\gamma_\sigma$ limit:
\begin{widetext}
\begin{equation}
n_{a,\mathrm f}^{(1)} \approx \Re\left\{ \frac{32g^2(\gamma_a^2\Omega^2 + 4i\gamma_a \Delta_a \Omega^2 - 4\Delta_a^2\Omega^2 - 8\Omega^4)\gamma_\sigma}{\gamma_a(\gamma_a+2i\Delta_a)^2(\gamma_a+2i\Delta_a-4i\Omega)^2(\gamma_a+2i\Delta_a
+4i\Omega)^2}\right\}.
\label{eq:n1filt}
\end{equation}
\end{widetext}
Interestingly, in the case of two-photon emission, this expression already appeared naturally in the formula we derived for $n_a^{(1)}$, since by substituting the two-photon condition  ${\Delta_a = \Delta_a^{(2)} = \Omega}$ in Eq.~\eqref{eq:n1filt} one obtains precisely the second term in the sum of Eq.~\eqref{eq:na-1photon-expansion}. Therefore, we can now understand such a term as the fraction of cavity population grown by first-order processes that is emitted at the resonant cavity frequency. By comparing this quantity to the population of $n$-photon bundles $n_a^{(n)}$, we can obtain an estimate of the purity of multiphoton emission for the light filtered at the cavity frequency and determine to which extent the figures of merit of multiphoton emission are improved.

\subsection{Figures of merit of the filtered emission.}

\begin{figure}[b!!]
\begin{center}
\includegraphics[width=0.95\columnwidth]{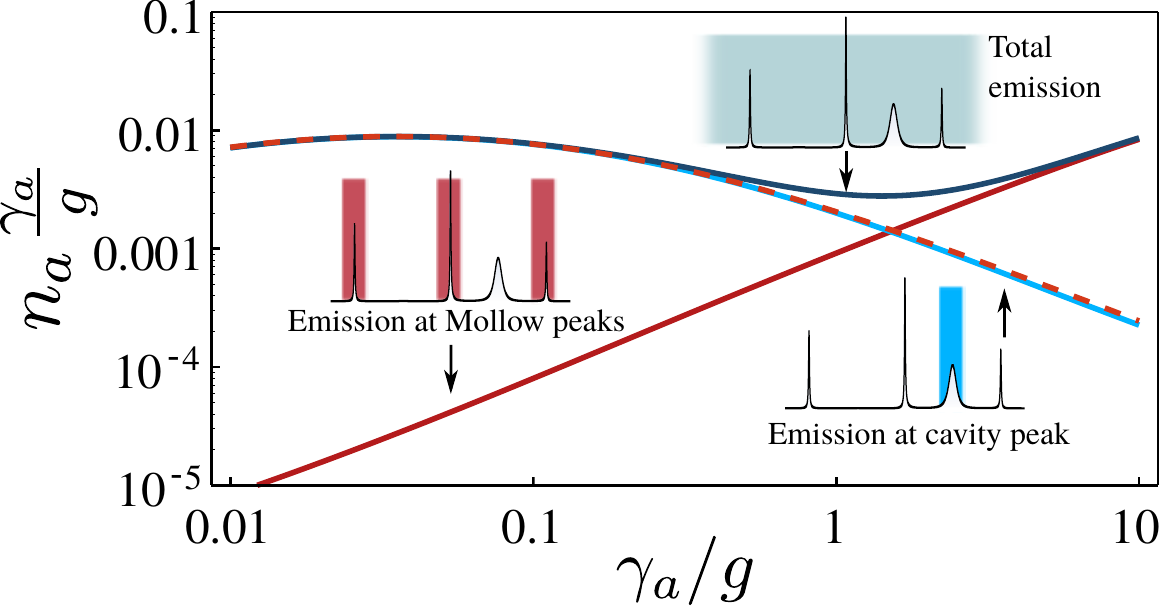}
\end{center}
\caption{Cavity emission rates at the two-photon resonance for three different spectral windows : central peak of the Mollow triplet (red), cavity peak (light blue) and total emission (dark blue). These rates have been numerically computed using Eq.~\eqref{eq:L}. The bundle emission $\gamma_a n_a^{(2)}$ (dashed, orange) computed from the $n$-photon master equation \eqref{eq:n-photon-ME} closely matches the emission at the cavity peak, confirming that the light emitted at that frequency is composed of photon bundles, and that spurious emission can be eliminated by frequency filtering.
Parameters: $\Omega/g=20$ and $\gamma_\sigma/g = 0.025$.}
\label{fig3-1}
\end{figure}
From the analysis made in the previous Section, it is expected that by filtering out the cavity emission one can improve substantially the figures of merit of our multiphoton source. This is based on the assumption that only part of the undesired photons---with a population corresponding to $n_{a,\mathrm{f}}^{(1)}$---will be emitted at the cavity frequency, whereas the totality of the $n$-photon bundles are emitted at such frequency. We can confirm this assumption by applying the previous analysis to the Liouvillian of the master equation in Eq.~\eqref{eq:n-photon-ME} as shown in Fig.~\ref{fig3-1}, where it is clearly seen how the two-photon population $n_a^{(2)}$---computed from Eq.~\eqref{eq:n-photon-ME}---corresponds to the amount of the total population $n_{a,\mathrm f}$ emitted at $\omega_a$. This is a particular case in which all the photons emitted at the cavity frequency are two-photon bundles and no spurious photons are emitted at that frequency. In a general situation this could fail to be the case, 
which leads us to the definition of the \emph{purity of filtered $n$-photon emission}, given by the fraction of the population emitted at the cavity frequency which is composed by $n$-photon bundles:
\begin{equation}
\pi_n^\mathrm{f} = \frac{n_a^{(n)}}{n_{a,\mathrm f}}\, ,
\end{equation}
which in the case of two-photon emission can be approximated as $\pi_2^\mathrm{f} \approx n_a^{(2)}/(n_{a,\mathrm f}^{(1)}+n_{a}^{(2)})$.
To illustrate the improvement brought by filtering, we show in Fig.~\ref{fig2}(b) a contour plot of the emission rates $n_{a}^{(2)}$ [red] and $n_{a,f}^{(1)}$ [blue] as a function of $(\gamma_a,\gamma_\sigma)$ with the same $\Omega=20g$, than the one used in Fig.~\ref{fig2}(a). By comparing the two panels, we observe that the region in which the two-photon filtered emission dominates over the single-photon one is substantially enlarged with respect to the unfiltered situation, even including regions in the weak-coupling regime $\gamma_a/g<1$. This is a major result, as it opens up the possibility of producing such multiphoton emission in systems within the bad-cavity limit, where $\gamma_{a,\sigma}<g$, but $C>1$, which is commonplace in both cavity and waveguide QED in the optical regime~\cite{goban13a,thompson13a,laucht12a,makhonin14a}.

\begin{figure}[b!!]
\begin{center}
\includegraphics[width=0.99\columnwidth]{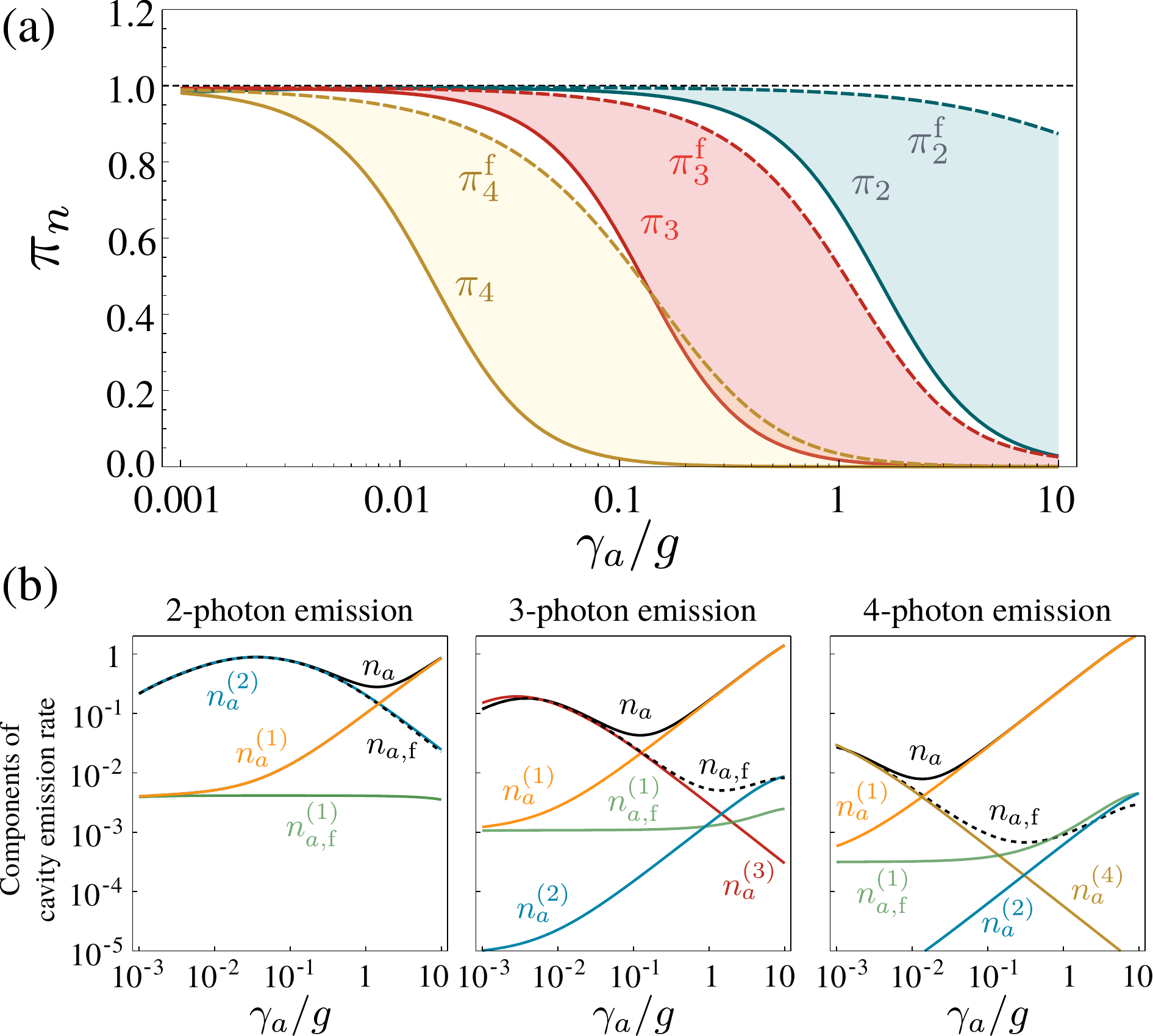}
\end{center}
\caption{(a) Numerical calculations for the purity of two, three and four-photon emission for unfiltered ($\pi_n$, solid) and filtered ($\pi_n^{f}$, dashed) cases, given by Eqs.~\eqref{eq:purity} and \eqref{eq:pi_f_general}, with $n_a^{(1)}$ and $n_{a,\mathrm f}^{(1)}$ computed numerically from a model truncated at one photon, and with $n_a^{(n)}$ for $n\geq 2$ computed numerically from the master equation \eqref{eq:n-photon-ME}. Parameters: $\Omega/g = 20$, $\gamma_\sigma/g = 0.025, 0.005$ and $0.001$ for $n=2,3,4$ respectively. (b) Numerically computed unfiltered and filtered emission rates, $\gamma_a n_a$ (solid, black) and $\gamma_a n_{a,\mathrm f}$ (dashed, black), compared to the underlying $n$-photon components of the cavity emission rate, $\gamma_a n_a^{(n)}$, used in the computation of $\pi_n$ and $\pi_n^\mathrm{f}$ of panel (a). }
\label{fig3}
\end{figure}
\begin{figure}[t!!!]
\begin{center}
\includegraphics[width=0.95\columnwidth]{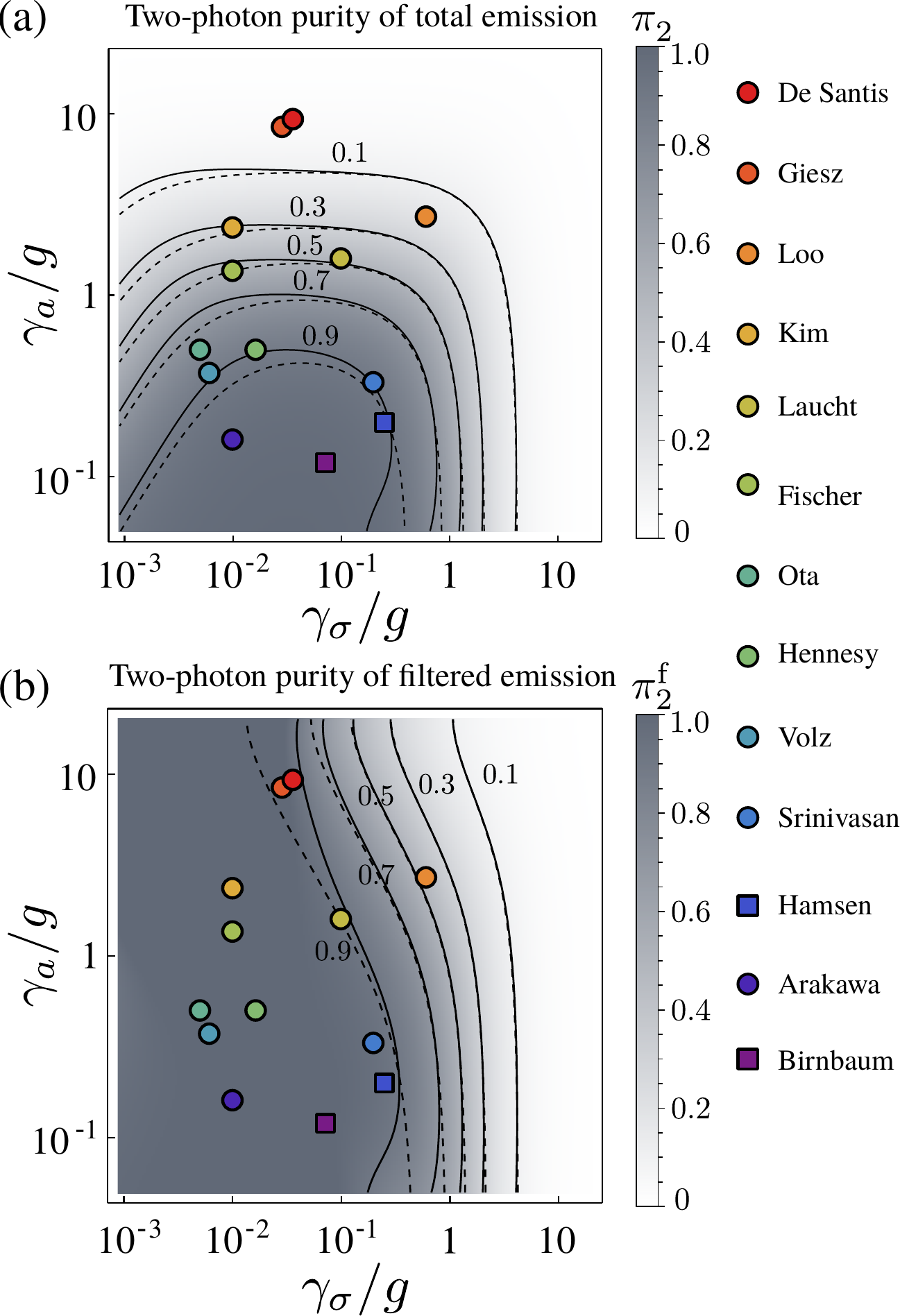}
\end{center}
\caption{Purity of two-photon emission in the unfiltered (a) and filtered (b) case, as a function of cavity decay rate $\gamma_a$ and 2LS decay rate $\gamma_\sigma$. Colored points correspond to experimental state of the art samples, with values summarized in Table~\ref{table:experiments}. Circles correspond to semiconductor QDs, and squares correspond to atoms. Parameters: $\Omega = 20g$. }
\label{fig3-2}
\end{figure}
\begin{table}
\begin{tabular}{l l r r r}
Reference & $\hbar g$ & $\gamma_a/g$ & $\gamma_\sigma/g$ & $C$\\
\hline\\
\textsc{Semiconductors} & &&&\\
De Santis~\emph{et al.} (2017)~\cite{desantis17a}& \SI{19}{\micro eV} & $9.4$ & $0.036$& $\sim 12$ \\
Giesz~\emph{et al.} (2016)~\cite{giesz16a} & \SI{21}{\micro eV} & $8.56$ & $0.028$ & $\sim 17$\\
Loo~\emph{et al.} (2012)~\cite{loo12a}& \SI{33}{\micro eV} & $2.76$ & $0.6$ & $\sim 2$ \\
Kim~\emph{et al.} (2014)~\cite{kim14a}& \SI{63}{\micro eV} & $2.35$ & $0.01$& $\sim 170$ \\
Laucht~\emph{et al.} (2009) ~\cite{laucht09a} & \SI{60}{\micro eV} & $1.6$ & $0.1$ & $\sim 25$\\
Fischer~\emph{et al.} (2016)~\cite{fischer16a}& \SI{45}{\micro eV} & 1.3 & $\sim 0.01$~\cite{muller15b}& $\sim 300 $ \\
Ota~\emph{et al.} (2011)~\cite{ota11a} & \SI{51}{\micro eV} & $0.5$ & $0.016$ & $\sim 500$\\
Hennessy~\emph{et al.} (2007)~\cite{hennessy07a} & \SI{90}{\micro eV} & $1.1$ & $\sim 0.005$  & $\sim 1600$\\
Volz~\emph{et al.} (2012)~\cite{volz12a} & \SI{141}{\micro eV} & $0.37$ & $\sim 0.006$ \footnote{Value not provided in the reference. We assumed the typical value of the QD lifetime in a photonic bandgap of $\sim$~\SI{5}{\ns}.}    & $\sim 1800$ \\
Srinivasan~\emph{et al.} (2007)~\cite{srinivasan07a} & \SI{12}{\micro eV} & $0.33$ & $0.2$ & $\sim 60$\vspace{5pt}
\\
Arakawa~\emph{et al.} (2012)~\cite{arakawa12a} & \SI{80}{\micro eV} & $0.3$ & $0.01$~\cite{ota11a} & $\sim 2500$\\
%
%
\textsc{Atoms} & &&&\\
Hamsen~\emph{et al.} (2016)~\cite{hamsen17a} & \SI{80}{\nano eV} & $0.2$ & $0.25$ & $\sim 80$ \vspace{5pt} \\
Birnbaum~\emph{et al.} (2005)~\cite{birnbaum05a} &\SI{0.14}{\micro eV} & $0.12$ & $0.071$ & $\sim 470$\\
\end{tabular}
\caption{Table of state-of-the-art parameters in cavity QED, with special emphasis on semiconductor samples. The high cooperativity values shown by semiconductor systems is because we are not considering in the definition the phonon-induced decoherence, that will reduce the effective cooperativity. We will consider its impact in detail in Section~\ref{subsec:phonon}.}
\label{table:experiments}
\end{table}

Using the analytical formulas we derived in the previous Section and Eq.~\eqref{eq:n1filt} for $n_{a,\mathrm f}^{(1)}$, we can obtain an asymptotic expression for the purity of {$n$-photon} filtered emission in the large driving limit, which for the two-photon case reads:
\begin{equation}
\pi_2^{\mathrm f} \approx \left(1 + \frac{8}{3C} + \frac{8 g^2}{\gamma_a^2 C^2} \right)^{-1} \, .
\label{eq:pi2-filt}
\end{equation}
By comparing this expression to Eq.\eqref{eq:pi2}, we can see clearly how frequency filtering allow us to get rid of the term proportional to $(\gamma_a/g)^2$ that enforces the strong coupling condition; now, the limit $\gamma_a \gg g$ leaves us an expression for $\pi_2^\mathrm{f}$ that depends only on the cooperativity, $\pi_2^{\mathrm f} \approx (1+\frac{8}{3C})^{-1}$. This provides a simple way to evaluate the maximum purity that can be observed for a system with a given cooperativity without the need to perform demanding numerical calculations.

For the general case of $n$-photon emission with $n>2$, the large driving limit of $\pi_n^\mathrm{f}$ goes as  $\pi_{n>2}^\mathrm{f}\propto \frac{1}{\Omega^{2(n-1)}}$. Even if this limit follows the same trend as in the unfiltered case, frequency filtering still provides a remarkable improvement in the figures of merit when an optimum driving is chosen. To illustrate it, we plot in Fig.~\ref{fig3}(a) the filtered and unfiltered purity for $n=2,3,4$ photons situation as a function of $\gamma_a/g$ for a system with $\Omega/g=20$ and $\gamma_\sigma/g=0.025, 0.005$ and $0.001$ respectively, where we observe a substantial improvement of the purity for all $n$. It is worth highlighting that 100\% of $n$-photon emission is guaranteed for good enough system parameters. Panel (b) depicts the underlying $n$-photon components of the cavity population, further corroborating our assumptions that the total population in the filtered and unfiltered case can, in good approximation, be decomposed into components associated to $n$-photon processes. This plot, however, also shows an effect not discussed so far: the resonant $n$-photon emission can also be spoiled by other off-resonant multiphoton processes of lower order $m<n$: in particular, we show that, in the case of three- and four-photon emission and for large values of $\gamma_a$, the population $n_a^{(2)}$ grown by the off-resonant two-photon process is larger than the corresponding resonant three- and four-photon population, and is the major factor in the emission at the cavity peak. This suggests a general expansion of $n_{a,\mathrm f}$ including all the spurious contributions to $n$-photon emission from off-resonant, lower $m$th-order processes (with $m<n$), so that the filtered purity can be approximated as:
\begin{equation}
\pi_n^\mathrm{f}  \approx \frac{n_a^{(n)}}{n_{1,\mathrm f}^{(1)}+\sum_{m=2}^{n}n_a^{(m)}}.
\label{eq:pi_f_general}
\end{equation}

Finally, to illustrate the feasibility of our proposal with state-of-the-art systems, we show in Fig.~\ref{fig3-2} the corresponding contour plots of both $\pi_2$ and $\pi_2^{\mathrm{f}}$ for the range of $(\gamma_a,\gamma_\sigma)$ displayed in Fig.~\ref{fig2}(a-b) and include points with several cavity QED experimental parameters (summarized in Table~\ref{table:experiments}), evidencing that many of them lie already within a regime of $\pi_2^{\mathrm{f}}\approx 1$. Note that we have adopted a definition of $\gamma_\sigma$ that only considers the decay to free space as the 2LS decoherence source neglecting, e.g., pure dephasing, which is usually relevant in semiconductor scenarios. In Section~\ref{sec:setups}, however, we consider the impact of these extra decoherence channels on our multiphoton emission.

\section{Impact of experimental limitations \label{sec:setups}}

Up to know, we have considered the ideal situation in which the only decoherence sources are both the cavity and 2LS losses, in which the cavity decay generally dominates, $\gamma_{a}\gg\gamma_\sigma$, and where the distinction between the cavity, laser and 2LS photons can be made perfect. Even though this ideal situation can be obtained within circuit QED setups~\cite{houck12a,you05a,arXiv_gu17a}, in the optical regime some of these requirements are more difficult to achieve.

The goal of this Section is to analyze the impact of several experimental imperfections in platforms relevant for the implementation of our proposal in the optical regime, such as semiconductor or atomic cavity QED setups. We focus on the case of two-photon emission, which is the one with more immediate prospects to be experimentally implemented. In particular, we analyze: i) the impact of phonon-induced decoherence in Section~\ref{subsec:phonon}; ii) the effect of detecting coherently scattered photons in the cavity emission in Section~\ref{subsec:imperfect}; and finally, iii) the robustness of the multiphoton emission in systems with limited driving and limited cancellation of spontaneous emission, that is, $\gamma_a\approx \gamma_\sigma$, in Section~\ref{subsec:limited}.

\subsection{Phonon-induced decoherence\label{subsec:phonon}}

One of the main problems in semiconductor cavity QED implementations is the decoherence produced by the phonons induced by the lattice vibrations in the solid. It is a well known fact that phonons in cavity QED systems~\cite{roy11a} give rise to two effects. The first one is the so-called pure dephasing mechanism, in which the phonons spoil the coherence of the 2LS without affecting directly the population. This mechanism can be described in a master equation description through the Lindblad term $\frac{\gamma_\phi}{2}\mathcal{L}_{\sigma^\dagger \sigma}[\rho]$. 

Moreover, when the 2LS is also interacting with a cavity mode, the phonons open an extra channel that allows to exchange excitations incoherently between the cavity and the 2LS. This is the \emph{cavity feeding} mechanism~\cite{hohenester09a,hohenester10a,roy11a,muller15a}, described by the following Lindblad terms:
\begin{align}
\frac{\gamma_{\sigma^\dagger a}}{2}\mathcal L_{\sigma^\dagger a}[\rho]+
\frac{\gamma_{\sigma a^\dagger}}{2}\mathcal L_{\sigma a^\dagger}[\rho],
\end{align}
where, following Ref.~\cite{roy11a}, we estimate the phonon transfer rates to be given by:
\begin{equation}
\gamma_{\sigma^\dagger a/\sigma a^\dagger} = 2\langle B\rangle ^2 g^2 \Re\left[\int_0^\infty d\tau\, e^{\pm i \Delta_a \tau } \left( e^{\phi(\tau)}-1 \right)\right],
\label{eq:phonon-rates}
\end{equation}
where $\langle B \rangle$ is:
\begin{equation}
\langle B \rangle = \exp\left[ -\frac{1}{2}\int_0^\infty d\omega \frac{J(\omega)}{\omega^2}\coth(\beta \hbar \omega/2) \right],
\end{equation}
$\phi(t)$ is:
\begin{equation}
\phi(t) = \int_0^\infty d\omega \frac{J(\omega)}{\omega^2}[\coth(\beta \hbar \omega/2)\cos(\omega t) - i\sin(\omega t)],
\end{equation}
and $J(\omega)$, the characteristic phonon spectral function, is defined as: $J(\omega) = \alpha_p \omega^3 \exp\left(-\frac{\omega^2}{2\omega_b^2} \right)$. Obviously, the larger the temperature is, the more relevant these mechanisms are.
\begin{figure}[t!!!]
\begin{center}
\includegraphics[width=0.99\columnwidth]{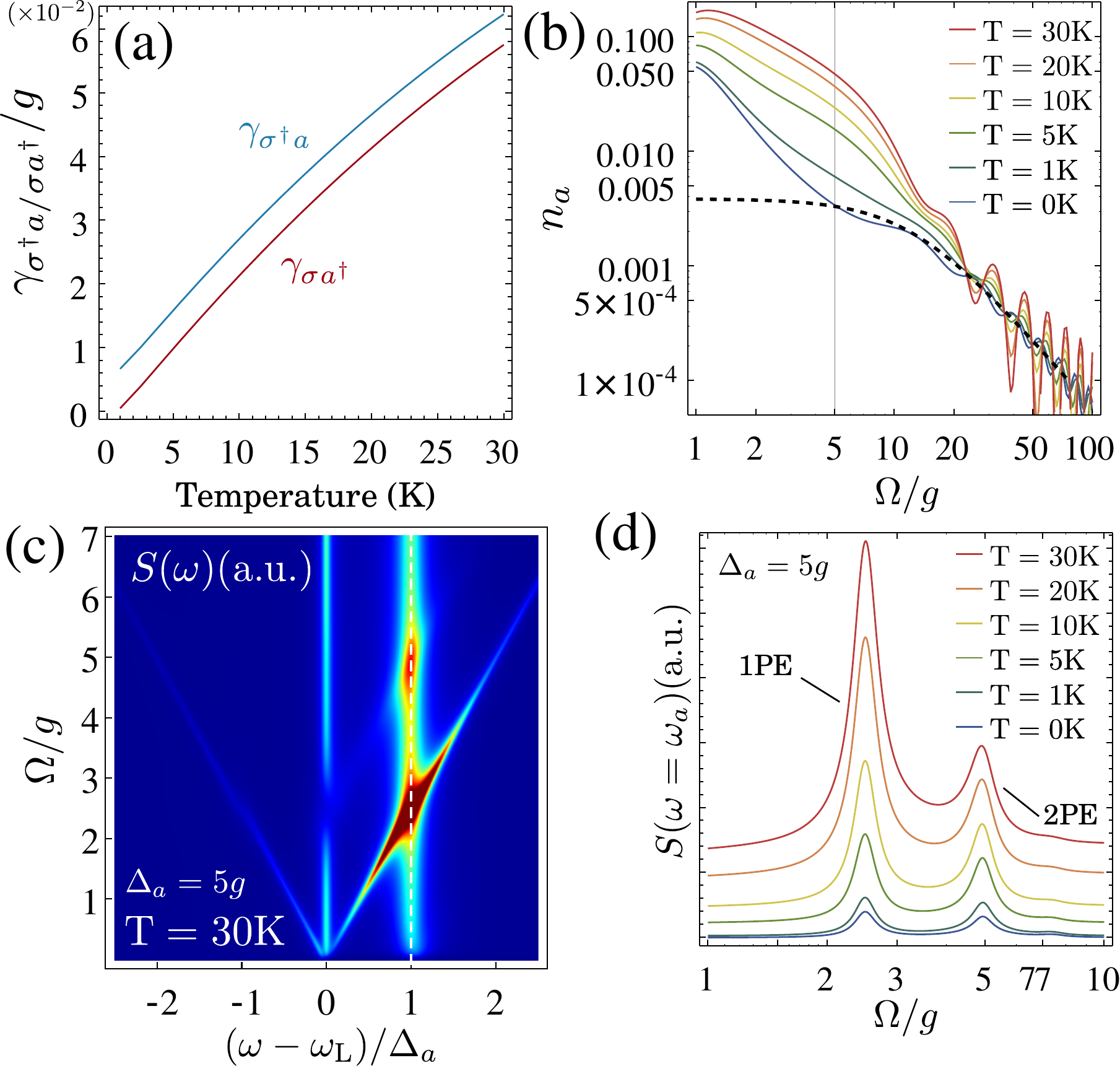}
\end{center}
\caption{(a) Calculated rates for the phonon-induced transitions as a function of temperature. (b) Cavity population emitted at the cavity frequency as a function of $\Omega$, for a cavity at the two photon resonance $\Delta_a = \Omega$ and different temperatures. Dashed, black: two-photo population.  (c) Spectrum of cavity emission as a function of the amplitude of the driving field for T = 30K. $\Delta_a = 5g$ (d) Spectrum at the cavity frequency as a function of driving field amplitude and different temperatures. $\Delta_a = 5g$. }
\label{fig4}
\end{figure}
In Fig.~\ref{fig4}, we illustrate the impact of these two mechanisms in the two-photon emission as a function of temperature. To do so, we choose a pure dephasing rate proportional to temperature, $\gamma_\phi=A\,T$, with $A=$\SI{1}{\micro eV/\K}~\cite{roy11a,bayer02a}, and set the parameters as $\omega_b = 0.22$~meV, and $\alpha_p = 0.18~\mathrm{meV}^{-2}$, consistently with the those provided in Refs.~\cite{muller15b,fischer16a}. The resulting values for $\gamma_{\sigma^\dagger a}$ and $\gamma_{\sigma a^\dagger}$ as a function of temperature are shown explicitly in Fig.~\ref{fig4}(a). In Fig.~\ref{fig4}(b) we study the effect of the temperature on $n_{a,\mathrm f}$, the cavity population emitted at the cavity frequency, in a plot similar to panel (c) of Fig.~\ref{fig2}: the filtered emission is shown as a function of $\Omega$, and compared with the amount of two-photon population, $n_a^{(2)}$. We observe that for drivings $\Omega/g\gtrsim 20$, the two-photon population still dominates the emission [$n_a^{(2)}\approx n_{a,\mathrm f}$], even for relatively large temperatures of $T\approx 30$ K. The oscillatory behaviour stems from the oscillatory dependence with $\Delta_a$ in Eq.~\eqref{eq:phonon-rates}, magnified by the logarithmic scale. At smaller driving amplitudes, the extra broadening introduced by phonons increase the emission of spurious single-photons, therefore decreasing the importance of the two-photon mechanism. 

It is interesting to highlight that even for the highest temperature of $T=30$ K, the spectral signature of the two-photon processes in $S(\omega)$ remains unambiguous, even for systems with moderate driving strength, as we show in the contour plot of Fig.~\ref{fig4}(c), where we plot the evolution of $S(\omega)$ as a function of the driving amplitude for a fixed $\Delta_a=5g$ as we did in Fig.~\ref{fig1}(a). There, we observe that apart from the expected broadening of the peaks in $S(\omega)$ and the enhancement of single-photon features, we still observe a well distinguished peak when $\Omega$ is such that $\Delta_a=\Delta_a^{(2)}=5g$. To make more evident that phonons do not preclude the observation of two-photon emission features we finally plot in Fig.~\ref{fig4}(d), the value of the cavity spectrum at the cavity frequency $\omega=\omega_a$ as a function of $\Omega$ and temperature $T$ and for a fixed $\Delta_a=5g$.  Remarkably, we observe that phonon-induced transitions not only results in an enhancement of single-photon emission, expected from the larger overlap with cavity peak, but also enhance the two-photon emission peak. This indicates that unambiguous signatures of two-photon physics can be observed even at high temperatures.

\subsection{Driving via the cavity mode: laser coherent back-scattering\label{subsec:imperfect}}

\begin{figure}[t!!!]
\begin{center}
\includegraphics[width=0.99\columnwidth]{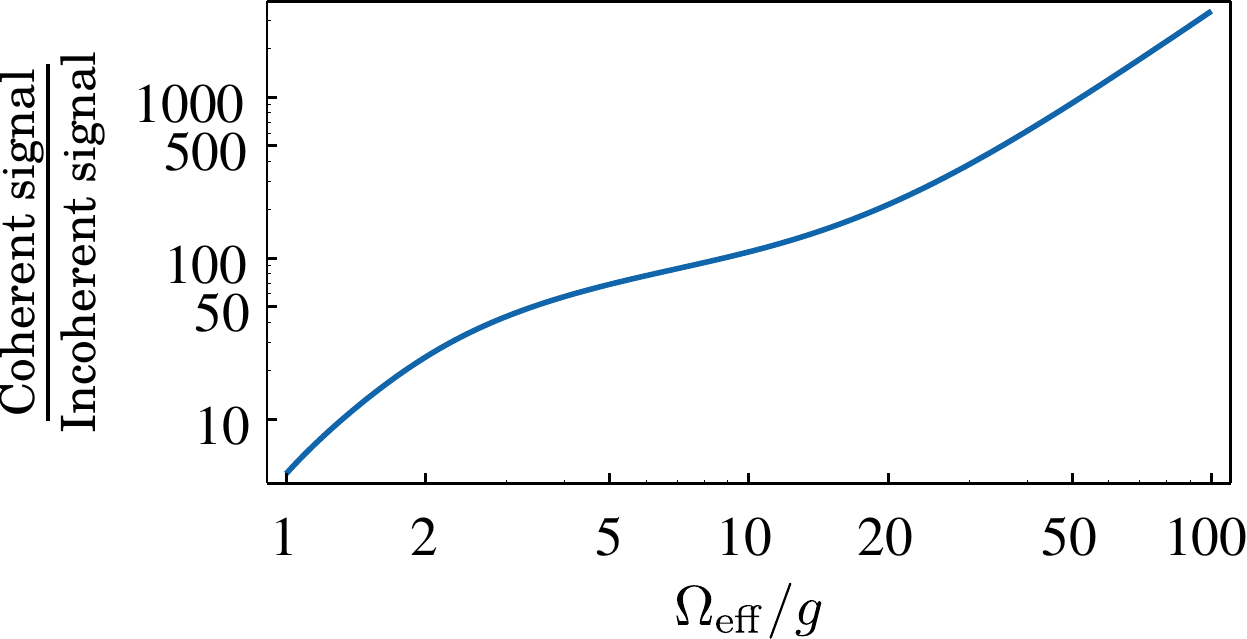}
\end{center}
\caption{Ratio between the coherent and incoherent signal at the cavity frequency when the cavity is coherently driven, as a function of the effective driving of the 2LS. In order to describe the situation where the full cavity peak is filtered, the filter linewidth has been taken equal to the cavity linewidth, $\Gamma = \gamma_a$. For each value of the driving, the cavity frequency is tuned to the two-photon resonance, $\Delta_a = \Omega_\mathrm{eff}$. }
\label{fig5}
\end{figure}

In our model we have assumed than the 2LS can be coherently driven independently of the cavity channel. Though this is certainly a possible configuration, in both semiconductor~\cite{kim14a,fischer16a} and atom cavity QED setups~\cite{hamsen17a}, the most common scenario experimentally is the one where the 2LS is driven coherently through the cavity channel. In order to describe that situation we replace $\Omega(\sigma+\sigma^\dagger)\rightarrow \Omega(a+a^\dagger)$, in the Hamiltonian of Eq.~\eqref{eq:Ham}. We can nevertheless come back to the description in terms of 2LS driving that we have used so far by rewriting the cavity operator as $a\rightarrow \alpha + a$, with $\alpha$ a complex number, and setting $\alpha = \Omega/(\Delta_a-i\gamma_a/2)$. This choice cancels the cavity driving terms in the master equation (where now the cavity operator describes the creation of quantum fluctuations on top of the coherent state built by the laser), and gives a Hamiltonian with an effective coherent driving in the 2LS given by $\Omega_{\mathrm{eff}}(e^{-i\phi}\sigma 
+ e^{i\phi}\sigma^\dagger)$, with $\phi = \arg{(\alpha)}$ and $\Omega_{\mathrm{eff}} = |\Omega/(\Delta_a-i\gamma_a/2)|$. This means that all the results obtained so far apply to this case by just replacing $\Omega$ by $\Omega_\mathrm{eff}$.

However, this type of driving leads to a potential problem, that is, the presence of coherently scattered light in the cavity spectrum, $S(\omega)=S_{\mathrm{I}}(\omega)+S_\mathrm{C}(\omega)$, which obviously spoils the multiphoton response and that needs to be rejected. The amount of coherently scattered photons $S_\mathrm{C}(\omega)$ when we place a filter of linewidth $\Gamma$ is given by:
\begin{equation}
S_\mathrm{C} = \frac{\Gamma^2}{2}\frac{|\langle a \rangle|^2}{\Gamma^2/4+\omega^2}\,.
\end{equation}
The ratio $S_\mathrm C(\omega_a)/S_\mathrm I(\omega_a)$ dictates what would be the experimental rejection of coherently scattered light required to avoid the laser background to overcome the incoherent signal. In Fig.~\ref{fig5}, we plot this ratio as a function of $\Omega_{\mathrm{eff}}$ in the case of two-photon emission, i.e., $\Delta_a = \Omega_\mathrm{eff}$, for a filter linewidth equal to the width of the cavity peak, $\Gamma =  \gamma_a$ and a cavity decay rate representative o semiconductor samples, $\gamma_a \approx g$. We see that, even for effective drivings as large as $\Omega_\mathrm{eff} \sim 100 g$, the ratio between coherent and incoherent scattered signal remains below $10^4$, which can be understood since in order to achieve two-photon emission the detuning between the cavity and the driving also increases with $\Omega_\mathrm{eff}$. This shows that, in that case, state-of-the-art rejection ratios of $\sim 10^{-6}$~\cite{kuhlman13a, flagg09a,he13a} would yield only $\sim 1$\% of detected photons coming from the coherent signal, and only $\sim 0.01$\% for drivings of $\Omega_\mathrm{eff} \approx 20$, which we have shown in previous sections to be sufficient to achieve regimes of perfect two-photon emission, $\pi_2^\mathrm{f}\approx 1$. The present analysis of the required rejection ration is also relevant in the context of recent developments that aim to achieve on-chip suppression of the coherent light by self-homodyned techniques~\cite{fischer16a,fischer17a}.

\begin{figure}[t!!!]
\begin{center}
\includegraphics[width=0.99\columnwidth]{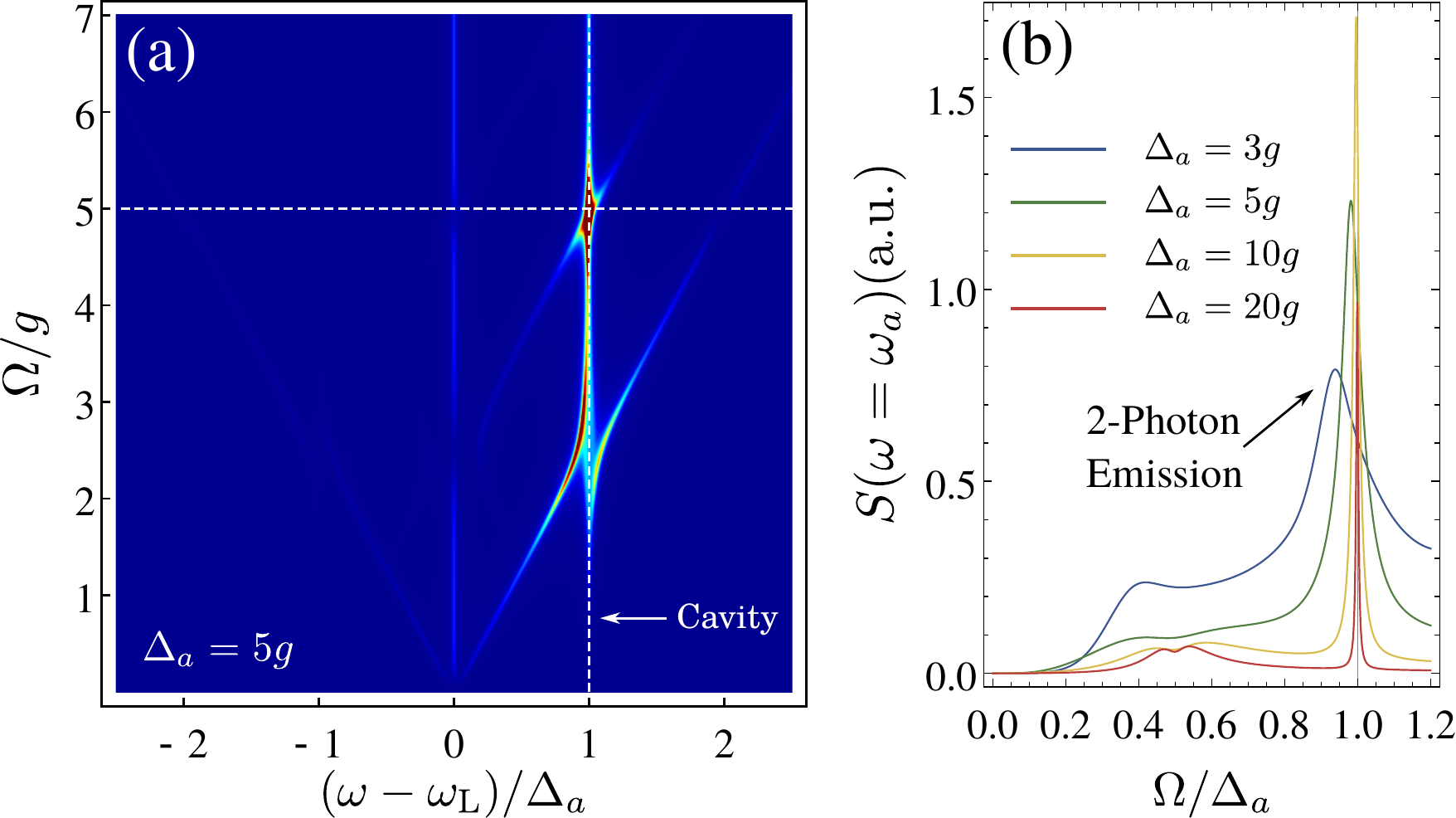}
\end{center}
\caption{(a) Cavity spectrum as a function of the driving amplitude $\Omega$ for set of parameters typical of atomic cavity QED systems, ${\gamma_a =\gamma_\sigma = 0.1g}$. The detuning between cavity and 2LS is $\Delta_a = 5g$. (b) Value of the spectrum at the cavity frequency $\omega_a$ as a function of the driving amplitude $\Omega$, for four fixed sets of cavity frequencies. A feature indicating two-photon emission appears whenever $\Omega = \Delta_a$, and it appears for driving amplitudes as low as $3g$. }
\label{fig6}
\end{figure}

\subsection{Limited driving amplitude \label{subsec:limited} and limited cancellation of spontaneous emission}

Finally, we also explore in more detail how the spectral signatures for multiphoton emission looks in the regime of parameters of atomic cavity QED systems~\cite{hamsen17a}. A typical restriction in those systems is that the 2LS spontaneous emission is not suppressed, such that it is $\gamma_\sigma\approx \gamma_a$, but with the advantage of being in the strong non-linear coupling $g>\gamma_{a,\sigma}$. Another limitation is that the drivings can not be very large, as it will decrease the trapping lifetime of the atom.  In order to illustrate that the two-photon emission signatures are still visible in this regime of parameters, we plot the evolution of $S(\omega)$ as a function of $\Omega$ for a fixed $\Delta=5g$ and a system with $\gamma_a=\gamma_\sigma=0.1g$. Comparing with Fig.~\ref{fig1}(a), we observe that: i) at the single-photon resonance, an anticrossing with the resonant sideband is present as the system is in the strong non-linear coupling; ii) more interesting for our manuscript, there is still a very strong signature when $\Omega$ 
matches the 2-photon resonance at $\Omega=\Delta_a$.

Finally, we illustrate the effect of smaller drivings in Fig.~\ref{fig6}(b), where we plot the amount of emission at the cavity resonance $S(\omega_a)$ as a function of $\Omega$ for several  values of $\Delta_a$. For large $\Delta_a$, e.g., $20g$, that correspond to large drivings, one observes a very sharp peak at the two-photon resonance. As one decreases $\Delta_a$, the height of the peak decrease and it gets broader. However, even for very small detunings $\Delta_a=3g$, the two-photon spectral resonance is still observed.

\section{Conclusions \& Outlook\label{sec:conclu}}

In this work we have shown how to optimally exploit the mechanism of $n$-photon emission introduced in Ref.~\cite{sanchezmunoz14a}, based on a coherently driven 2LS coupled to a cavity mode. In particular, we have provided an analytical understanding of the figures of merit of the emission such as the efficiency rates, purity of $n$-photon emission and its spectral distribution in the cavity output field. Thanks to these closed-form expressions, we provide formulas for the optimal driving, $\Omega$, and the emitter's decay rate, $\gamma_\sigma$, to maximize the $n$-photon emission. More importantly, we identify that the spurious single photons can be filtered out, relaxing in some situations the strong coupling requirement used in the previous proposal, to the more accessible \emph{bad-cavity} regime, thus opening the possibility of implementing this protocol in a wide variety of platforms. 
Our results also allow us to foresee the future of this field in systems with parameters still out of reach for most platforms, that we have shown to produce close to 100\% three- and four-photon emission with demanding but still realistic parameters.
Finally, we analyse the impact of several experimental imperfections, such as phonon-induced decoherence, on the $n$-photon emission showing how they are within the reach of current experimental technologies in semiconductor and atom cavity QED setups.

In future works it will be interesting to introduce another controllable decay channel on the 2LS, e.g., another cavity mode, which allows one to control dynamically the decay rate of $\gamma_\sigma$. With that knob, one can tune the optimal $n$-photon emission regime, or control the separation between the $n$-photon states. Moreover, by measuring the emission of that channel, one can herald the presence of the $n$-photon states in the cavity one, as these processes can be shown to be strongly correlated~\cite{sanchezmunoz14a}. Other heralding protocols could benefit from exploiting higher-order frequency-resolved correlations in the emission~\cite{arXiv_lopezcarreno17a}. Further exciting perspectives are the triggered generation of $n$-photon states with appropriate pulse shaping of $\Omega(t)$, or the study of the interplay of these mechanism taking into account other spectral densities as the ones appearing in band-edges of waveguide QED setups.

\section{Acknowledgements}
The authors thank Kevin Fischer for stimulating discussions. C.S.M. acknowledges support by the RIKEN iTHES Project,
MURI Center for Dynamic Magneto-Optics via the AFOSR Award No. FA9550-14-1-0040,
the Japan Society for the Promotion of Science (KAKENHI),
the IMPACT program of JST,
JSPS-RFBR grant No 17-52-50023,
CREST grant No. JPMJCR1676,
and the John Templeton Foundation. F.P.L. acknowledges the POLAFLOW ERC project No. 308136. E.dV. acknowledges the project FIS2015-64951-R (CLAQUE) and RyC program. C.T. acknowledges financial support from the Spanish MINECO under contract MAT2014-53119-C2-1-R.  A.G.T. acknowledges support from Intra-European Marie-Curie Fellowship NanoQuIS (625955).

\appendix
\section{Derivation of the $n$-photon Hamiltonian}
\label{appendix:n-photon}
For the sake of generality, in this demonstration we consider the detuned situation $\Delta \equiv \omega_\sigma - \omega_\mathrm{L} \neq 0$, in which the eigenstates of the dressed 2LS-laser system are given by:
\begin{subequations}
\begin{align}
&|+\rangle = c|g\rangle + s|e\rangle\,,\\
&|-\rangle = s|g\rangle - c|e\rangle\, ,
\label{eq:c6-eigenvectors-mollow}
\end{align}
\end{subequations}
with
\begin{subequations}
\begin{align}
&c = 1/\sqrt{{1+\xi^{-2}}} \, ,\\
&s =1/\sqrt{1+\xi^2}\, ,\\
&\xi = \frac{\Omega}{\Delta/2 + R}\, ,\\
&R = \sqrt{\Omega^2 + \left(\frac{\Delta}{2} \right)^2}\, .
\end{align}
\end{subequations}

When the cavity is resonant with the $n$-th photon transition, we consider the subspace spanned by the states $\{|+,0\rangle, |-,n\rangle, |+,1\rangle, |-,1\rangle,\cdots,|+,n-1\rangle |-,n-1\rangle  \}$, where the Hamiltonian reads:
\begin{equation}
\mathcal{H}^{(n)} = 
\begin{pmatrix}
\hat h^{(n)} & \hat{V}^{(n)}\\
{\hat V^{(n) \mathrm{T}}} & \hat H^{(n)}
\end{pmatrix}\, . 
\end{equation}
In this way, we have divided the Hamiltonian in two parts, with $\hat h^{(n)}$ acting on the subspace $\{|+,0\rangle, |-,n\rangle \}$:
\begin{equation}
\hat h^{(n)} = \begin{pmatrix} R & 0 \\ 0 & -R + n \Delta_a \end{pmatrix},
\end{equation}
and $\hat H^{(n)}$ acting on the rest of our global subspace. For the lowest multiphoton resonance, $n=2$, $\hat H^{(2)}$  is defined as
\begin{equation}
\hat H ^{(2)} = \begin{pmatrix}
R + \Delta_a & 0\\
0 & -R + \Delta_a
\end{pmatrix},
\end{equation}
from which we can build recursively the general $n>2$ case as:
\begin{equation}
\hat H ^{(n)} = \begin{pmatrix} 
\hat H^{(n-1)} & \hat{X}^{(n) \mathrm{T}}\\
\hat{X}^{(n)} & \begin{matrix}
R + (n-1)\Delta_a & 0\\
0 & -R + (n-1)\Delta_a
 \end{matrix}
 \end{pmatrix},
\end{equation}
where
\begin{equation}
 \hat X^{(n)} = 
 \begin{pmatrix}
0 & 0 & \sqrt{n-1} g c s & -\sqrt{n-1} g c^2\\
0 & 0 & \sqrt{n-1} g s^2 & -\sqrt{n-1} g c s 
  \end{pmatrix}.
\end{equation}
The two subspaces on which $\hat h^{(n)}$ and $\hat H^{(n)}$ act are coupled by $\hat V^{(n)}$, a $2\times 2(n-1)$ matrix given by:
\begin{equation}
\hat V^{(n)} =  \begin{pmatrix} g c s & g s^2 & \cdots & 0 & 0 \\ 
0& 0& \cdots & \sqrt{n} g c s^2 & -\sqrt{n} g c s\end{pmatrix} \, .
\end{equation}
with no zeros in the ${n=2}$ case. Since the sign of $\Delta_a$ is irrelevant for the physics, we will consider it to be positive and set to the $n$-photon resonance, $\Delta_a \approx 2R/n$. Our purpose now is to obtain an effective Hamiltonian $\hat h^{(n)}_\mathrm{eff}$ within the subspace $\{|+,0\rangle,|-,n\rangle \}$ by means of matrix perturbation theory:
\begin{equation}
\hat h_\mathrm{eff}^{(n)} = \hat h^{(n)} + \hat V^{(n)}(E_0 - \hat H^{(n)})^{-1} \hat V^{(n) \mathrm{T}}.
\label{eq:c6-heff}
\end{equation}
This calculation is easy to perform and allows to obtain analytic expressions for $\hat h^{(n)}_\mathrm{eff}$ that, however, are too lengthy to be written here. Nevertheless, by inspecting the result of computing Eq.~\eqref{eq:c6-heff} up to $n=6$, we obtained the off-diagonal term of $\hat h_\mathrm{eff}^{(n)}$ to lowest order in $g$, given by $\hat h_{\mathrm{eff}\, 1,2} = \sqrt{n} g^{(n)}$, where $g^{(n)}$ is an effective $n$-photon coupling rate given by:
\begin{equation}
g^{(n)} = \frac{g^n}{R^{n-1}}\left(\frac{n^2}{2}\right)^{(n-1)}\frac{c^{n-1}s^{n+1}}{(n-1)!^2} +\mathcal O(g^{n+1}) \, .
\label{eq:c6-gn}
\end{equation}
which for the case of $\Delta = 0$ takes the form of Eq.~\eqref{eq:gn}.

\section{Derivation of the expressions of $n_a^{(n)}$ and $n_a^{(1)}$}
\label{appendix:na_n}
The derivation of the steady-state $n$-photon population $n_a^{(n)}$ predicted by the master equation Eq.\eqref{eq:n-photon-ME} is estimated by writing the following closed set of differential equations for the correlators:
\begin{subequations}
\begin{align}
\frac{d}{dt}\langle a^n\tilde\sigma^\dagger \rangle  &\approx  \left(2iR-n i\Delta_a-\frac{n\gamma_a + P + \gamma_{\tilde\sigma} + \gamma_{\tilde\phi}}{2}\right) \langle a^n\tilde\sigma^\dagger\rangle \, , \nonumber\\
&-i n! g^{(n)} \langle \tilde\sigma^\dagger \tilde\sigma \rangle \, , \\
\frac{d}{dt}\langle \tilde\sigma^\dagger \tilde\sigma  \rangle &= -(\gamma_{\tilde\sigma}+P)\langle \tilde\sigma^\dagger \tilde\sigma \rangle + 2 g^{(n)} \Im\left\{\langle a^n\tilde\sigma^\dagger\rangle \right\} \nonumber\\ &+P\, ,\\
\frac{d}{dt}\langle a^\dagger a \rangle &= -\gamma_a \langle a^\dagger a \rangle - 2ng^{(n)} \Im\left\{\langle a^n\tilde\sigma^\dagger\rangle \right\} \, .
\end{align}
\end{subequations}
where the missing correlators were neglected under the assumption $g^{(n)} \ll \gamma_a$. By setting these derivatives to zero, one obtains the expression of $n_a^{(n)}$ given in Eq.~\eqref{eq:na-nphoton}.

For the estimation of $n_a^{(1)}$ we consider the master equation in Eq.~\eqref{eq:master-equation} with the Hamitlonian of Eq.~\eqref{eq:Ham}, and write the differential equations for the correlators truncating at one cavity photon. This yields the following set of equations:

\begin{subequations}
\begin{align}
\frac{d\mean{a^\dagger a}}{dt} =& -\gamma_a \mean{a^\dagger a} - 2 g\Im{\mean{{{\sigma}}^\dagger a}}\, ,\\
\frac{d\langle {\sigma}^\dagger \sigma \rangle}{dt} =& -\gamma_{{\sigma}} \mean{{\sigma}^\dagger \sigma} - 2\Omega \Im{\mean{\sigma}}-2g\Im{\mean{a^\dagger {{\sigma}}}},
\label{eq:c6-SDE-n-sigma}
\\
\frac{d\mean{{{\sigma}}}}{dt} =& -\left(i\Delta+\frac{\gamma_{{\sigma}}}{2}\right)\mean{{{\sigma}}} + 2i\Omega \mean{{\sigma}^\dagger \sigma} - ig\mean{a} \nonumber \\ &+ 2ig\mean{a {\sigma}^\dagger \sigma} - i\Omega,
\label{eq:c6-SDE-sigma}
\\
\frac{d\mean{a}}{dt} =& -\left(i\Delta_a+\frac{\gamma_a}{2} \right)\mean{a} - ig\mean{{{\sigma}}},
\label{eq:c6-SDE-a}\\
\frac{d\mean{{{\sigma}}^\dagger a}}{dt} =& -\left[ i(\Delta+ \Delta_a) + \frac{\gamma_{{\sigma}}+\gamma_a}{2} \right]\mean{{{\sigma}}^\dagger a} - 2i\Omega \mean{a {\sigma}^\dagger \sigma} \nonumber \\
&+i\Omega \mean{a} -2ig\mean{a^\dagger a {\sigma}^\dagger \sigma}-ig\mean{{\sigma}^\dagger \sigma} + ig\mean{a^\dagger a},
\label{eq:c6-SDE-sigma-dagger-a}\\
\frac{d\mean{a {\sigma}^\dagger \sigma}}{dt} =& -\left(i\Delta_a+\frac{\gamma_a}{2} + \gamma_{{\sigma}} \right)\mean{a {\sigma}^\dagger \sigma} + ig\mean{a^\dagger a {{\sigma}}} \nonumber \\ & -i\Omega \mean{a{{\sigma}}^\dagger} +i\Omega\mean{a{{\sigma}}} \, ,
\label{eq:c6-SDE-a-n-sigma}
\\
\frac{d\mean{a{{\sigma}}}}{dt} =& -\left[i(\Delta+\Delta_a)+\frac{\gamma_{{\sigma}} + \gamma_a}{2} \right]\mean{a{{\sigma}}}-i\Omega \mean{a} \nonumber \\ &+  2i\Omega \mean{a {\sigma}^\dagger \sigma} \, .
\end{align}

\end{subequations}

Under the approximation  $\Omega \gg g,\gamma_a,\gamma_\sigma$, we can eliminate the terms proportional to $g\mean{a^\dagger \tilde\sigma}$ in Eq.~\eqref{eq:c6-SDE-n-sigma}, to $ig\mean{a}$ and $ig\mean{a{\tilde\sigma}^\dagger\tilde\sigma}$ in Eq.~\eqref{eq:c6-SDE-sigma}, to $g\mean{a^\dagger a {\tilde\sigma}^\dagger \tilde \sigma}$ and $g\mean{a^\dagger a}$ in Eq.~\eqref{eq:c6-SDE-sigma-dagger-a} and to $g\mean{a^\dagger a \tilde \sigma}$ in Eq.~\eqref{eq:c6-SDE-a-n-sigma}. This leaves us, setting the derivatives to zero, with the following set of equations for the steady state observables:

\begin{subequations}
\begin{align}
\mean{a^\dagger a}_{\mathrm SS} &= -\frac{1}{\gamma_a } 2 g\Im{\mean{{\sigma}^\dagger a}_{\mathrm SS}}\, ,\\
\label{eq:nsigma}
\mean{{\sigma}^\dagger  \sigma}_{\mathrm SS} &\approx  -\frac{1}{\gamma_{\sigma}}2\Omega \Im{\mean{\sigma}_{\mathrm SS}} \, ,\\
\label{eq:sigma}
\mean{\sigma}_{\mathrm SS} &\approx  \frac{i\Omega}{i\Delta+\frac{\gamma_{\sigma}}{2}}(2\mean{{\sigma}^\dagger  \sigma}_{\mathrm SS}-1)\, ,\\
\label{eq:a}
\mean{a}_{\mathrm SS} &\approx -\frac{g}{\Delta_a-i\frac{\gamma_a}{2}}\mean{\sigma}_{\mathrm SS} \, , \\
\label{eq:sigmapa}
\mean{{\sigma}^\dagger a}_{\mathrm SS} &\approx  i\frac{-2\Omega\mean{a {\sigma}^\dagger\sigma}_{\mathrm SS} + \Omega \mean{a}_{\mathrm SS} - g\mean{{\sigma}^\dagger\sigma}_{\mathrm SS}}{i(\Delta+\Delta_a)+\frac{\gamma_a+\gamma_{\sigma}}{2} }\, ,\\
\label{eq:ansigma}
\mean{a {\sigma}^\dagger\sigma}_{\mathrm SS} &\approx  \frac{i\Omega(\mean{a\sigma}_{\mathrm SS}-\mean{{\sigma}^\dagger a}_{\mathrm SS})}{\frac{\gamma_a}{2}+\gamma_{\sigma} + i\Delta_a}\, ,\\
\mean{a{\sigma}}_{\mathrm SS} &= \frac{-i\Omega(2\mean{a {\sigma}^\dagger \sigma}_{\mathrm SS}-\mean{a}_{\mathrm SS})}{i(\Delta+\Delta_a)+\frac{\gamma_a+\gamma_{\sigma}}{2} }\, .
\end{align}
\end{subequations}

This is a closed system of equations that can be readily solved. From it, we can obtain the population ${n_a^{(1)} \equiv \mean{a^\dagger a}_\mathrm{SS}}$, which for the case $\Delta=0$ reads:
\begin{multline}
n_{a}^{(1)} = -\frac{16 g^3}{\gamma_a } \Im\left\{ \phantom{\frac{16 g^3}{\gamma_a }} \right. \\ \left. 
\frac
{\Omega^2\left[-i8\Omega^2(\gamma_a+2i\Delta_a)   -i\lambda(\Delta_a)^2\lambda_2(\Delta_a)\right]}
{(\gamma_a+2i\Delta_a)\lambda(\Delta_a)(8\Omega^2+\gamma_{\sigma}^2)(\lambda(\Delta_a)\lambda_2(\Delta_a)+16\Omega^2)}
\right\}\, ,
   \label{eq:c6-na-1photon}
\end{multline}
where
 $\lambda(\Delta_a) \equiv \gamma_a + \gamma_{\sigma}+2 i \Delta_a$ and
 $\lambda_2(\Delta_a) \equiv \gamma_a + 2\gamma_{\sigma}+2 i \Delta_a$.
 In the particular case of two-photon resonance condition, $\Delta_a = \Delta^{(2)} = \Omega$, the expansion of Eq.~\eqref{eq:c6-na-1photon} to first order in $\gamma_\sigma/g$ yields Eq.~\eqref{eq:na-1photon-expansion}.

\section{Derivation of $n_{a,\mathrm{f}}^{(1)}$}
\label{appendix:n1-filt} 

In this section we get an analytical estimation of the amount of light emitted at the cavity peak in a system where the cavity's Hilbert space is truncated at one photon, thus suppressing multiphoton processes. To do so, it is convenient to calculate the spectrum using the quantum regression theorem for the evolution of a vector of correlators. Let us assume that the dynamics of the mean value $\mean{a}$ is coupled to a set of correlators that we write in a vector $\mathbf{u}$, ${a}$ being the first element, $\mathbf{u}_1 = {a}$. In our case, the vector is $\mathbf{u} = ({a},{\sigma},{\sigma^\dagger},{\sigma^\dagger \sigma})^\mathrm{T}$, and the dynamics of the mean values follows the equation:
\begin{equation}
\frac{d}{dt}\mean{\mathbf{u}} = 
M \mean{\mathbf{u}} + \mathbf{c}\, ,
\label{eq:c6-first-QRT-edo}
\end{equation}
with 
\begin{subequations}
\begin{align}
 M =& \begin{pmatrix}
-\frac{\gamma_a}{2}-i\Delta_a & -ig & 0 & 0\\
-i g & -\frac{\gamma_{\sigma}}{2}-i\Delta & 0 & 2i\Omega\\
0 & 0& -\frac{\gamma_{\sigma}}{2} + i\Delta & -2i\Omega\\
0 & i\Omega & -i\Omega & -i\gamma_{\sigma}
\end{pmatrix} \, ,\\
\mathbf{c} =& \begin{pmatrix}0 & -i\Omega & i\Omega & 0 \end{pmatrix}^\mathrm{T}\,,
\end{align}
\end{subequations}
where we assumed $\Omega \gg \gamma_\sigma, \gamma_a, g$. 
In order to obtain an homogeneous equation, we define a vector $\mathbf{v}$ given by ${\mathbf{v} = \mathbf{u}- \mathbf{u}_\mathrm{SS}}$, where $\mathbf{u}_\mathrm{SS}$  are the steady state solutions of Eq.~\eqref{eq:c6-first-QRT-edo}, giving the equation:
\begin{equation}
\frac{d}{dt}\mean{\mathbf{v}} = 
M \mean{\mathbf{v}} \, .
\label{eq:c6-second-QRT-edo}
\end{equation}

We can now apply the quantum regression theorem and write the dynamics of the set of two-time correlators:
\begin{equation}
\frac{d}{d\tau}\mean{a^\dagger(t)\mathbf{v}(t+\tau)}  = 
M \mean{a^\dagger(t)\mathbf{v}(t+\tau)}.
\label{eq:QRT}
\end{equation}
Defining $\mathbf{w}(\tau) = \mean{a^\dagger(t)\mathbf{v}(t+\tau)}$, the spectrum is then given by:
\begin{equation}
S(\omega) = \frac{1}{\pi}\Re\int_0^\infty \mathbf{w}_1(\tau)e^{i\omega \tau} d\tau.
\label{eq:c6-spectrum-QRT}
\end{equation}
The vector $\mathbf{w}(\tau)$ has the formal solution  $\mathbf{w}(\tau) = e^{M\tau}\mathbf{w}(0)$, which, considering the diagonalized form of $M$:
\begin{equation}
-D = E^{-1}M E
\end{equation}
where $E$ is the matrix of normalized eigenvectors of $M$, reads $\mathbf{w}(\tau) = E e^{-D\tau}E^{-1}\mathbf{w}(0)$. This allow us to integrate formally Eq.~\eqref{eq:c6-spectrum-QRT} and find an expression for the spectrum analogous to Eq.~\eqref{eq:c6-spectrum-lorentzians}, with $D_\beta = \gamma_\beta/2 + i\omega_\beta$,  and:
\begin{equation}
L_\beta \equiv \Re\{E_{1,\beta}[E^{-1}\mathbf{w}(0)]_\beta \}.
\end{equation}
Since we worked with an approximated closed system of four correlators, we have only four eigenvalues $D_\beta$, and therefore the spectrum is composed of four Lorentzian peaks. One of these eigenvalues, that we assign to $\beta=1$, has an imaginary part equal to $\omega_a$, and therefore corresponds to the Lorentzian peak of the spectrum at the cavity frequency. From it, we can define the part of the background population of photons that are emitted at the cavity frequency as:
\begin{equation}
n_{a,\mathrm f}^{(1)} \equiv L_1\, .
\end{equation}
This corresponds to the number of background photons that cannot be spectrally separated, since they are emitted at the cavity frequency. Thanks to the small size of the matrix $M$, we can obtain approximate analytical expressions for $n_{a,\mathrm f}^{(1)}$. Expanding to first order in $\gamma_\sigma/g$, we obtain the expression given in Eq.~\eqref{eq:n1filt}.

\bibliography{Sci-Alex,Sci-Carlos,books,arXiv}

\end{document}